\documentclass[sigconf]{acmart}
\usepackage{multirow}
 \usepackage{enumitem} 
 \usepackage{adjustbox}
\usepackage[ruled,vlined]{algorithm2e}
\usepackage{subcaption}
\usepackage{array}

\AtBeginDocument{%
  }

\setcopyright{acmlicensed}
\copyrightyear{2026}
\acmYear{2026}
\acmDOI{XXXXXXX.XXXXXXX}
\acmConference[Conference acronym 'XX]{Make sure to enter the correct
  conference title from your rights confirmation email}{June 03--05,
  2018}{Woodstock, NY}

\acmISBN{978-1-4503-XXXX-X/2018/06}

\begin{document}

\title{Disagreement as Signals: Dual-view Calibration for Sequential Recommendation Denoising}

\author{Sijia Li}
\affiliation{%
  \institution{School of Big Data and Software Engineering, Chongqing University}
  \country{China}
}
\email{lisijia@stu.cqu.edu.cn}

\author{Min Gao}
\affiliation{%
  \institution{School of Big Data and Software Engineering, Chongqing University}
  \country{China}
}
\email{mingao@cqu.edu.cn}

\author{Zongwei Wang}
\affiliation{%
  \institution{School of Big Data and Software Engineering, Chongqing University}
  \country{China}
}
\email{zongwei@cqu.edu.cn}

\author{Zhiyi Liu}
\affiliation{%
  \institution{School of Big Data and Software Engineering, Chongqing University}
  \country{China}
}
\email{zhiyi@cqu.edu.cn}

\author{Xin Xia}
\affiliation{%
  \institution{University of Queensland}
  \country{Australia}
}
\email{xin.xia@uq.edu.au}

\author{Yi Zhang}
\affiliation{%
  \institution{University of Queensland}
  \country{Australia}
}
\email{yi.zhang@uq.edu.au}

\renewcommand{\shortauthors}{Trovato et al.}

\begin{abstract}

Sequential recommendation seeks to model the evolution of user interests by capturing temporal user intent and item-level transition patterns. Transformer-based recommenders demonstrate a strong capacity for learning long-range and interpretable dependencies, yet remain vulnerable to behavioral noise that is misaligned with users’ true preferences. Recent large language model (LLM)–based approaches attempt to denoise interaction histories through \textbf{static} semantic editing. Such methods neglect the learning dynamics of recommendation models and fail to account for the evolving nature of user interests.
To address this limitation, we propose a \underline{D}ual-view 
\underline{C}alibration framework for \underline{S}equential
\underline{R}ecommendation denoising(DC4SR). Specifically, we introduce a \textit{semantic prior}, derived from an LLM fine-tuned via labeled historical interactions, to estimate the noise distribution from a semantic perspective. From the learning perspective, we further employ a \textit{model-side posterior} that infers the noise distribution based on the model’s learning dynamics.
The disagreement between the two distributions is then leveraged to jointly refine semantic understanding and learning-aware model-side representations. Through iterative updates, \textbf{dynamic} dual-view calibration is achieved for both the global semantic prior and the model-side posterior, enabling consistent alignment with evolving user interests. Extensive experiments demonstrate that DC4SR consistently outperforms strong Transformer-based recommenders and LLM-based denoising methods, exhibiting enhanced robustness across training stages and noise conditions. Our implementation is available at \url{https://anonymous.4open.science/r/DC4SR-4892/}.
\end{abstract}



\maketitle

\section{Introduction}
Recommender systems (RecSys) have demonstrated substantial business value across a wide range of applications, including e-commerce and online media platforms~\cite{ko2022survey, schafer2001commerce}. 
As an important branch of RecSys, the core objective of RecSys is to predict the next item that a user is likely to engage with based on historical interactions~\cite{wang2019sequential}. 
To better align recommendations with real-world user behavior, sequential recommendation has been introduced to explicitly model the temporal evolution of user interests and capture item-level transition patterns from interaction sequences~\cite{wang2019sequential,kang2018self,sun2019bert4rec}. Among existing approaches, Transformer-based sequential recommenders have become the dominant paradigm~\cite{kang2018self,sun2019bert4rec}, as they transform user histories from compressed hidden states into explicit and interpretable dependency structures, enabling effective modeling of long-range behavioral dependencies~\cite{vaswani2017attention}.

\begin{figure}[t]
  \centering
  \includegraphics[width=\linewidth]{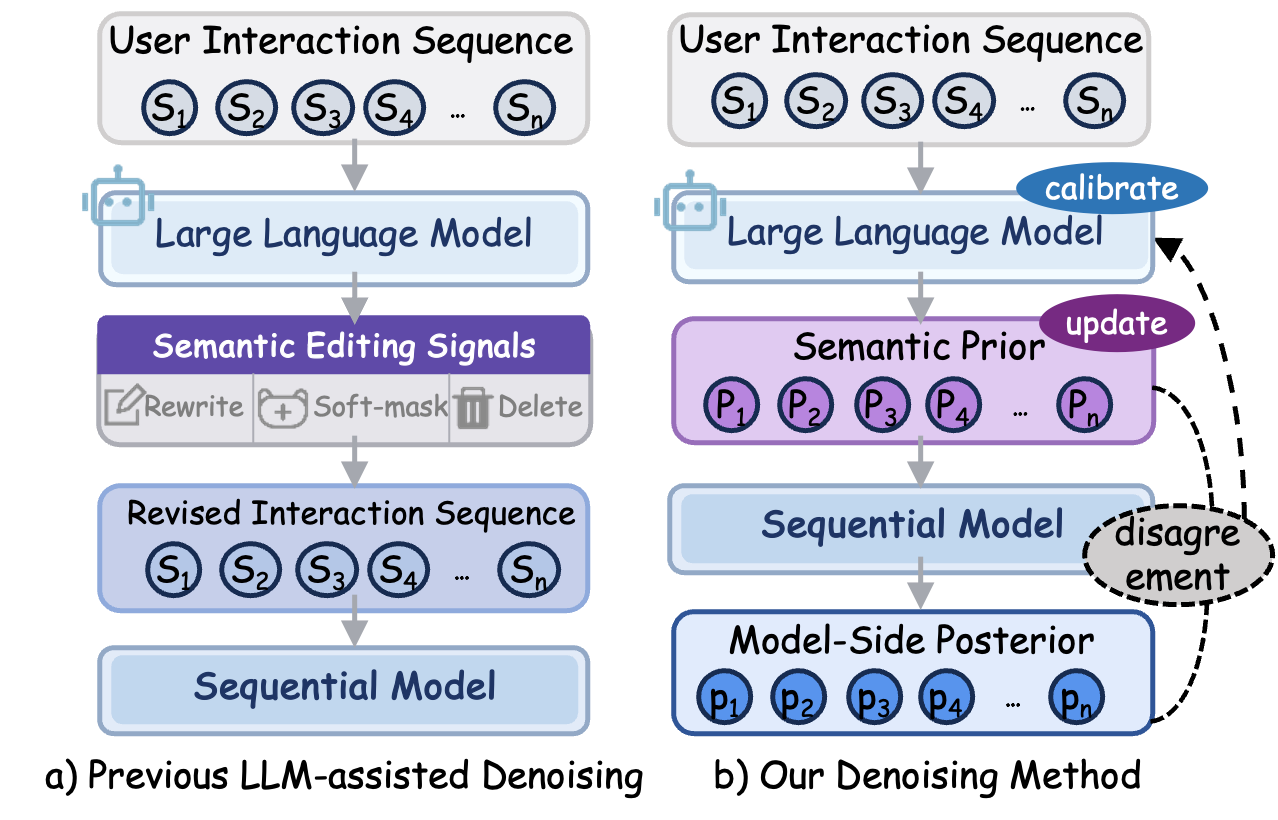}
  \caption{LLM-assisted denoising paradigms. (a) Static semantic editing. (b) Dual-view calibration between the semantic prior the and model-side posterior.}
  \label{fig:llm_denoising_paradigm}
\end{figure}

Despite the effectiveness of Transformer-based recommenders, real-world interaction sequences are inherently contaminated by behavioral noise arising from accidental clicks, transient interest shifts, and external interference such as advertisements~\cite{joachims2007evaluating, o2006detecting, martinez2016managing, wang2021clicks}. For Transformer-based recommenders, such noise is particularly problematic, as the self-attention mechanism tends to propagate such noise across the entire sequence~\cite{kang2018self,sun2019bert4rec}. As a result, noisy interactions can induce a \textbf{global} deviation from the user’s true intent while simultaneously causing \textbf{temporal} misalignment with the user’s evolving interests~\cite{martinez2016managing, han2024efficient}. To mitigate this issue, conventional denoising methods identify suspicious interactions using heuristic rules~\cite{wang2022learning}, self-supervised or contrastive objectives~\cite{lin2023self,xie2022contrastive}, or representation-based criteria~\cite{zhang2022hierarchical}, and subsequently remove, replace, or reweight these interactions during training. While these approaches are effective at filtering salient noise, their reliance on surface-level patterns and implicit assumptions limits their ability to uncover subtle yet pervasive noise in user histories.

\begin{figure}[t]
  \centering
  \includegraphics[width=\linewidth]{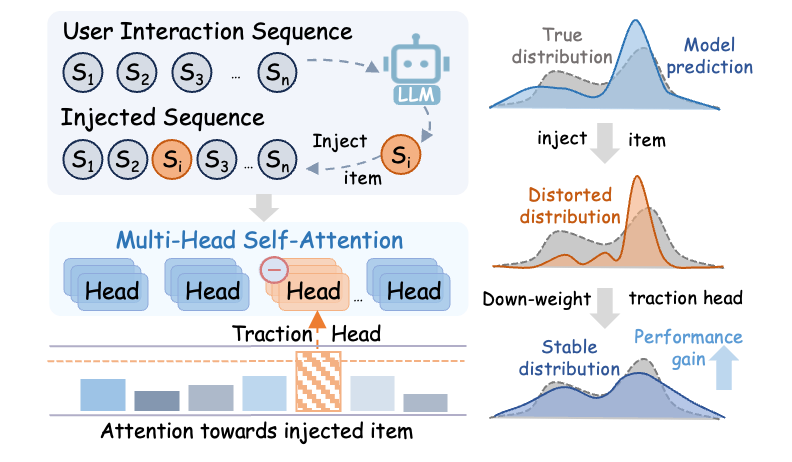}
  \caption{Traction analysis illustration.}
  \label{fig:traction_illustration}
\end{figure}

Recent large language model (LLM)–based methods attempt to overcome this limitation by leveraging the strong semantic understanding and commonsense reasoning capabilities of LLMs~\cite{zhang2023denoising, liu2025large, wu2024survey}. These approaches typically extract semantic signals from item descriptions and interaction contexts to identify noisy behaviors as a pre-processing step, followed by \textbf{static} sequence rewriting, interaction removal, or soft reweighting~\cite{wang2025llm4dsr, wu2025empowering, sun2025llm4rsr}, as illustrated in the left of Figure~\ref{fig:llm_denoising_paradigm}. Although such methods improve interpretability and noise discovery, their effectiveness fundamentally relies on a \textbf{global} semantic view of the interaction sequence, which remains incomplete~\cite{zhang2025shapley, ji2023survey}. In particular, these static pre-processing steps fail to account for the evolving influence of noise throughout the learning process, leading to misalignment with model-side \textbf{temporal} learning dynamics~\cite{zhang2025shapley}. In contrast, as we demonstrate in Section~\ref{sec:traction} and Figure~\ref{fig:traction_illustration}, sequential models exhibit pronounced sensitivity to noise during training, \textit{i.e.}, a small subset of attention heads responds abnormally to perturbations, even when injected items are semantically similar to the original sequence. This observation suggests that model-side learning dynamics encode noise-related signals beyond static semantic similarity, highlighting the need for learning-aware denoising mechanisms that effectively integrate both global and temporal views.

Motivated by these observations, we move beyond static semantic denoising and propose DC4SR,  a \underline{D}ual-view \underline{C}alibration framework for \underline{S}equential
\underline{R}ecommendation denoising that jointly leverages semantic and model-side assessments throughout learning. As illustrated in Figure~\ref{fig:llm_denoising_paradigm}, DC4SR effectively maintains two complementary views and explicitly exploits their disagreement as a calibration signal. The LLM-based semantic view provides a \textbf{global prior}, derived from an LLM fine-tuned via labeled historical interactions, over interaction reliability, while the model-side view infers a \textbf{temporal posterior} from model-side training signals, aligned with the sequential recommender’s evolving learning dynamics. When substantial disagreement emerges at specific positions, these interactions are more likely to be ambiguous or training-critical and are therefore \textbf{dynamically} revisited rather than irrevocably filtered by semantic decisions. On the model side, DC4SR further incorporates a head-aware modulation mechanism that adaptively suppresses attention heads exhibiting abnormal traction to suspicious interactions. Guided by the identified training-critical positions, the semantic prior is periodically recalibrated by refining the LLM on the most unreliable interactions, enabling semantic denoising to co-evolve with the backbone model. Through this dual-view, disagreement-driven calibration process, DC4SR achieves robust and adaptive denoising that consistently aligns sequential representations with users’ evolving interests.


Our main contributions are summarized as follows:
\begin{itemize}
    \item We propose DC4SR, a \underline{D}ual-view \underline{C}alibration framework for \underline{S}equential \underline{R}ecommendation denoising that jointly leverages semantic and model-side assessments, and interprets their disagreement to dynamically calibrate noisy interactions during training.
    \item We perform traction analysis to identify noise-sensitive heads and enable head-level regularization, and propose a budgeted periodic refresh that updates the semantic prior on unreliable sequences and contentious positions, addressing global and temporal noise effects.
    \item We conduct systematic experiments across multiple datasets and noise settings, showing consistent improvements over sequential and LLM-assisted denoising baselines, and include a disagreement analysis that supports the effectiveness and complementarity of the dual-view calibration.
\end{itemize}

\section{Preliminaries}
\subsection{Sequential Recommendation}
We follow the standard setting of sequential recommendation.
Let $\mathcal{U}$ and $\mathcal{I}$ denote the user and item sets.
For each user $u\in\mathcal{U}$, we observe a chronological interaction sequence
$S_u^{1:n}(s_{u,1},\ldots,s_{u,n})$ with $s_{u,n}\in\mathcal{I}$.
Given a prefix $S_u^{1:n}$, the goal is to predict the next item $s_{u,n+1}$.

We adopt a Transformer-based sequential recommender.
Each item $i$ is associated with an embedding $\mathbf{e}_i\in\mathbb{R}^d$ and each position $t$ with a positional embedding $\mathbf{p}_t\in\mathbb{R}^d$,
where $d$ is the embedding dimension.
For a prefix of length $t$, the input at position $\tau\in\{1,\ldots,t\}$ is
$\mathbf{h}_\tau^{(0)}=\mathbf{e}_{i_{u,\tau}}+\mathbf{p}_\tau$,
and a stack of $L$ Transformer blocks produces the final hidden state $\mathbf{h}_t^{(L)}$ at the last position,
where $L$ is the number of blocks.
We score a candidate item $i\in\mathcal{I}$ by
\begin{equation}
\hat{y}_{u,t}(i)=\mathbf{h}_t^{(L)\top}\mathbf{e}_i .
\end{equation}
With the positive item $i^+=i_{u,t+1}$ and a sampled negative set $\mathcal{N}_{u,t}$, we optimize a BCE objective
\begin{equation}
\mathcal{L}_{\mathrm{BCE}}=-\sum_{(u,t)}\left[\log\sigma(\hat{y}_{u,t}(i^+))+\sum_{j\in\mathcal{N}_{u,t}}\log\!\big(1-\sigma(\hat{y}_{u,t}(j))\big)\right],
\end{equation}
where $\sigma(\cdot)$ is the sigmoid function.

\subsection{LLM Fine-Tuning for Semantic Prior}
\label{sec:prelim_llm_finetune}
We fine-tune an LLM with LoRA\cite{hu2022lora} on our instruction-formatted supervision data to improve its ability to identify noise-like interactions from item texts, which will be used to derive a semantic signal in Sec.~\ref{sec:method_semantic_prior}.
Each supervision instance is an instruction--response pair: the prompt serializes a user prefix with item texts, and the response provides a textual noise judgment conditioned on the prefix context and item texts.
Table~\ref{tab:llm_sft_format} summarizes the data fields and the prompt template.

We adopt fine-tuning with LoRA and keep the base model weights frozen.
Let $\theta$ denote the frozen base parameters and $\phi$ denote the trainable LoRA adapter parameters; the resulting model is denoted as $\mathcal{M}_{\theta,\phi}$, and we use $P_{\theta,\phi}(\cdot)$ for its token probabilities.
For a linear projection $W\in\mathbb{R}^{d_{\mathrm{out}}\times d_{\mathrm{in}}}$s($d_{\mathrm{in}}$ and $d_{\mathrm{out}}$ denote the input and output dimensions of an LLM projection layer) in the LLM, LoRA parameterizes the effective weight as
\begin{equation}
W' = W + \Delta W,
\qquad
\Delta W = BA,
\end{equation}
where $W$ is frozen, while $A\in\mathbb{R}^{r\times d_{\mathrm{in}}}$ and $B\in\mathbb{R}^{d_{\mathrm{out}}\times r}$ are trainable low-rank factors with rank $r\ll \min(d_{\mathrm{in}},d_{\mathrm{out}})$.

Given a prompt $p_i$ formed by concatenating \texttt{instruction} and \texttt{input}, and a target response $y_i$ from \texttt{output}, we fine-tune the LLM by minimizing the standard supervised fine-tuning objective over the response tokens:
\begin{equation}
\mathcal{L}_{\mathrm{SFT}}
=
-\sum_{i}\sum_{k=1}^{|y_i|}
\log P_{\theta,\phi}\!\left(y_{i,k}\mid p_i, y_{i,<k}\right),
\end{equation}
where the loss is computed only on the output part.

\begin{table}[t]
\centering
\caption{Instruction tuning prompt used for fine-tuning the LLM to identify noisy interactions, where the response provides a textual noise judgment.}
\small
\setlength{\tabcolsep}{6pt}
\renewcommand{\arraystretch}{1.10}
\begin{tabular}{p{0.18\linewidth} p{0.7\linewidth}}
\hline
Field & Template \\
\hline

\texttt{instruction} &
{\ttfamily
You are a recommender systems researcher familiar with Amazon-style user behavior logs and item title semantics.
Your task is to identify noise items in the user's behavior sequence based on the surrounding context and item texts.
Return the set of items you consider suspicious noise. Output noise items.
} \\

\texttt{input} &
{\ttfamily
User behavior sequence (chronological):\newline
1. <item-title/text> | 2. <item-title/text> \newline
... | T. <item-title/text>
} \\

\texttt{output} &
{\ttfamily
Suspicious items:\newline
\{ <item-title>, <item-title>, ... \}\newline
If none:\; \{\}
} \\

\hline
\end{tabular}
\label{tab:llm_sft_format}
\end{table}

\subsection{Head-Level Sensitivity}
\label{sec:traction}
To examine whether robustness degradation can be adequately explained purely by sequence semantics, we conduct a traction analysis on a transformer-based sequential recommender.
For each user, we employ an LLM to select a semantically related item and inject it into the interaction sequence while preserving the maximum length by truncating the oldest interaction.
As illustrated in Figure~\ref{fig:traction_illustration}, although the injected item is semantically plausible, it can trigger a disproportionate change in attention weights in a small subset of attention heads (traction heads), distorting the model prediction distribution.

We quantify head-level injection sensitivity using the traction gain $\Delta A_{l,h}$ and identify traction heads based on this score; experimental details and further conclusions are deferred to Appendix~\ref{app:traction}.
The resulting head-level sensitivity is highly non-uniform and concentrated in a few heads.
This structural sensitivity indicates that semantic plausibility alone does not determine whether an interaction is harmful at a given training stage, motivating model-side signals as training-time supervision for calibrating noisy interactions.

\section{Methodology}
\subsection{Overview}
\begin{figure*}[t]
  \centering
  \includegraphics[width=\textwidth]{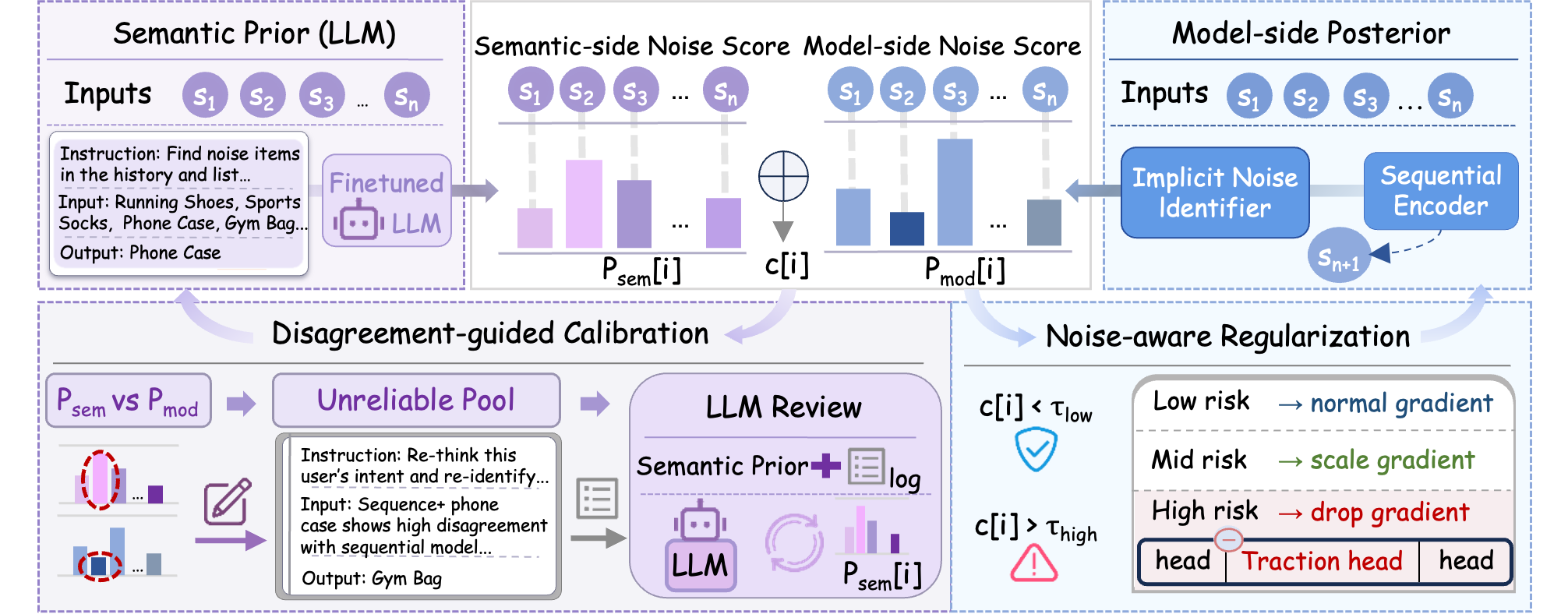}
  \caption{Overview of DC4SR. The LLM produces a position-wise semantic prior $p_{\mathrm{sem}}[i]$, the sequential model estimates a model-side posterior risk $p_{\mathrm{mod}}[i]$, and their disagreement $d[i]$ is fused into a training-time weight $c[i]$ to (i) trigger prior refresh on unreliable positions and (ii) gate noise-aware optimization.}
  \label{fig:framework}
\end{figure*}

As illustrated in Figure~\ref{fig:framework}, DC4SR performs dynamic dual-view calibration by combining an LLM-based semantic prior with model-side training signals for denoising sequential recommendation, aiming to reduce the global deviation and temporal misalignment induced by noisy interactions.
During training, the disagreement between the two views identifies contentious positions and triggers periodic, budgeted refresh of the semantic prior on localized hard cases.
Meanwhile, the fused risk weight drives noise-aware optimization, including position-wise gated gradient control and traction-informed attention-head regularization that suppresses heads with consistently high traction scores.

\subsection{LLM-Based Semantic Prior}
\label{sec:method_semantic_prior}

We use the instruction-tuned LLM from Sec.~\ref{sec:prelim_llm_finetune} as a lightweight semantic scorer to estimate a position-wise noise prior for historical interactions.
Instead of decoding free-form item lists, we obtain a vectorized prior by reading the next-token probability at a fixed response prefix, which avoids generation-time variance and can be computed with a single forward pass.

For a user history $S_u=(s_{u,1},\ldots,s_{u,n})$, we construct an inference prompt $x_u$ following Table~\ref{tab:llm_sft_format} and instantiate the output header as the fixed prefix \texttt{Suspicious items:}.
Let $z_i$ denote the anchor token used to represent item $s_{u,i}$ under the LLM tokenizer, and let $m_i\in\{0,1\}$ indicate whether such a single-token anchor exists.
We then read the next-token distribution at the first decoding step immediately following \texttt{Suspicious items:} and define the raw semantic noise score as
\begin{equation}
\ell_i \triangleq m_i \cdot \log P_{\theta,\phi}\!\left(z_i \mid x_u\right),
\quad i=1,\ldots,n.
\end{equation}
where $m_i=0$ assigns a zero score when no single-token anchor is available for $s_{u,i}$, and a larger $\ell_i$ indicates a higher semantic likelihood that position $i$ corresponds to a noise interaction under the context $x_u$.

To obtain a bounded prior in $(0,1)$, we directly map the raw score via a sigmoid:
\begin{equation}
p_{\mathrm{sem}}[i] = \sigma(\ell_i ),\quad i=1,\ldots,n,
\label{eq:p_sem}
\end{equation}
where $p_{\mathrm{sem}}[i]$ estimates the semantic noise probability for the $i$-th interaction.
The resulting vector $\mathbf{p}_{\mathrm{sem}}$ serves as the position-wise semantic prior and is fused with model-side evidence in subsequent modules.

\subsection{Model-Side Posterior}
\label{subsec:model_posterior}

The semantic prior $p_{\mathrm{sem}}[i]$ is derived from item texts and indicates whether the interaction at position $i$ appears noise-like in semantics. However, as shown in Section~\ref{sec:traction}, a transformer-based sequential recommender may assign overly concentrated influence to a small subset of positions through self-attention, making some interactions negatively affect ranking even when they look semantically plausible. Therefore, we introduce a model-side risk score to capture training-dependent shortcut reliance.

Given a user prefix $S_u^{1:n}$, the sequential encoder first produces hidden states
$\mathbf{h}_1,\ldots,\mathbf{h}_n$ with $\mathbf{h}_i\in\mathbb{R}^d$ and outputs the backbone prediction for each position.
Based on the encoder outputs, we compute a training-dependent evidence $\xi_i$ for each position using a training-time metric that reflects its risk under the current model state.
We then combine $\mathbf{h}_i$ and $\xi_i$ with an implicit noise identifier $\psi(\cdot)$ and obtain:
\begin{equation}
p_{\mathrm{mod}}[i]
=
\sigma\!\big(\psi(\mathbf{h}_i, \xi_i)\big)\in(0,1),
\quad i=1,\ldots,n.
\label{eq:model_posterior}
\end{equation}
We emphasize that $p_{\mathrm{mod}}$ is not a supervised noise label; it is a training-dependent proxy learned jointly with the backbone under the training objective.
Through this coupling, $\psi(\cdot)$ is encouraged to assign higher risk to positions that consistently attract large attention mass yet worsen the next-item likelihood, so the backbone is discouraged from concentrating on such keys.

The semantic prior and the model-side posterior describe risk from two complementary views. We quantify their mismatch by a position-wise disagreement score
\begin{equation}
d[i]=\big|p_{\mathrm{sem}}[i]-p_{\mathrm{mod}}[i]\big|,
\quad i=1,\ldots,n.
\label{eq:align_discrepancy}
\end{equation}
A large $d[i]$ indicates conflicting evidence between semantic-only and model-state risk. We use $d[i]$ as a hard case indicator to select suspicious positions for targeted refresh and calibration.

We next combine the semantic prior and the model-side posterior into a single training risk weight that reconciles the two evidence sources.
Concretely, we define
\begin{equation}
c[i]
=
\mathrm{clip}\Big(
\alpha\,p_{\mathrm{mod}}[i]
+(1-\alpha)\,p_{\mathrm{sem}}[i],\ 0,\ 1
\Big)\in[0,1],
\label{eq:fused_clip}
\end{equation}
where the linear form offers a simple and stable fusion of the two risk signals, and $\mathrm{clip}(\cdot)$ enforces the valid range for subsequent regularization.
Importantly, we keep disagreement separate from the fusion weight and do not introduce an additional conflict term into $c[i]$. Instead, the disagreement score $d[i]$ is used to flag hard positions where the two signals conflict.

We then regularize the backbone with a position-wise risk gate for decision-level modulation and a continuous risk-weighted penalty for attention allocation.
We further introduce two thresholds $\tau_{\mathrm{low}}$ and $\tau_{\mathrm{high}}$ with $0\le \tau_{\mathrm{low}} < \tau_{\mathrm{high}} \le 1$ to categorize positions into low-, mid-, and high-risk regimes. Concretely, we define a risk gate $g(\cdot)$ that modulates the contribution of a historical position:
\begin{equation}
g(c[i])=
\begin{cases}
1, & c[i] < \tau_{\mathrm{low}} \quad \text{(low risk)} ,\\
\eta_\theta, & \tau_{\mathrm{low}} \le c[i] \le \tau_{\mathrm{high}} \quad \text{(mid risk)} ,\\
0, & c[i] > \tau_{\mathrm{high}} \quad \text{(high risk)} ,
\end{cases}
\label{eq:risk_gate}
\end{equation}
where $\eta_\theta\in(0,1)$ is a learnable mid-risk attenuation scalar optimized with the backbone.
We set $\tau_{\mathrm{high}}=\mathrm{Quantile}_{0.9}(c)$ and $\tau_{\mathrm{low}}=\mathrm{Quantile}_{0.2}(c)$ on the training set, so that $g(c[i])=0$ for the highest-risk 10\% positions and $g(c[i])=1$ for the lowest-risk 20\% positions.

Finally, we use the fused risk weight $c[i]$ to derive head-level suppression signals from high-risk keys.
Let $a_h(t,k)$ denote the attention weight of head $h$ from query position $t$ to key position $k$.
We measure how much a head focuses on the high-risk set by
\begin{equation}
r_h=\mathbb{E}_{t}\!\left[\sum_{k\le t} a_h(t,k)\,\mathbb{I}\!\left[c[k]>\tau_{\mathrm{high}}\right]\right],
\label{eq:head_risk}
\end{equation}
and map it to a head-wise down-weighting coefficient $w_h\in(0,1]$ by normalizing $\{r_h\}$ across heads. A larger $r_h$ yields a smaller $w_h$.

We then penalize risk-weighted attention mass with head suppression by:
\begin{equation}
\mathcal{L}_{\mathrm{risk\text{-}attn}}
=
\mathbb{E}_{h,t}\!\left[
(1-w_h)\sum_{k\le t} a_h(t,k)\,c[k]
\right].
\label{eq:risk_attn}
\end{equation}
Together with $g(\cdot)$, $\mathcal{L}_{\mathrm{risk\text{-}attn}}$ acts as the continuous, head-aware component that penalizes attention mass on risky positions.

\subsection{Disagreement-Guided Calibration}
\label{subsec:disagreement_log}
The semantic prior provides a global signal about whether an interaction is suspicious given item semantics, while the model-side posterior reflects the backbone’s optimization-time signals at a dynamic training stage.
When the two signals agree, the interaction is consistently assessed under both views.
In contrast, a large disagreement between semantic and sequential signals indicates that the backbone assigns non-trivial training influence to an interaction that is not semantically salient, or does not emphasize an interaction that is flagged as suspicious by the semantic prior.
Such conflict points can be indicative of shortcut reliance, where self-attention may propagate spurious influence, making disagreement a targeted cue for review and prior refresh.

Based on this intuition, we perform disagreement-guided calibration with a budgeted periodic refresh schedule.
Every $R$ epochs (totalling $\lfloor E/R \rfloor$ rounds over $E$ epochs), we update the semantic prior only on the top-$M$ unreliable sequences and their top-$K$ most contentious positions.

Given valid history positions $\mathcal{V}(x)$ of a training sequence $x$, we define its sequence-level score as the average position-wise disagreement $d_x[i]$(Eq.~\ref{eq:align_discrepancy}):
\begin{equation}
U(x)=\frac{1}{|\mathcal{V}(x)|}\sum_{i\in\mathcal{V}(x)} d_x[i].
\label{eq:unreliability}
\end{equation}

In each refresh round, we construct an unreliable pool by selecting the top-$M$ sequences with the largest $U(x)$:
\begin{equation}
\mathcal{P}=\mathrm{TopM}\big(\{U(x)\}_{x\in\mathcal{D}}\big).
\label{eq:unreliable_pool}
\end{equation}
For each selected sequence $x\in\mathcal{P}$, we further localize hard cases by taking the top-$K$ positions with the largest $d_x[i]$:
\begin{equation}
\mathcal{H}(x)=\mathrm{TopK}\big(\{d_x[i]\}_{i\in\mathcal{V}(x)}\big).
\label{eq:hard_spots}
\end{equation}
This yields a compact set of sequences and positions with the highest semantic and model disagreement under the dual-view signals.

To avoid re-analyzing the whole sequence from scratch at every refresh, we maintain a sequence-level log $\mathcal{E}(x)$ for each $x\in\mathcal{P}$, which stores position-aligned traces accumulated across rounds.
At round $r$, for each hard case $i\in\mathcal{H}(x)$, we record an evidence entry
\begin{equation}
e^{(r)}_{x,i}=\Big(i,\ \mathcal{T}(s_{x,i}),\ p_{\mathrm{sem}}^{(r)}[i],\ p_{\mathrm{mod}}^{(r)}[i],\ d_x^{(r)}[i]\Big),
\label{eq:evidence_entry}
\end{equation}
where $\mathcal{T}(s_{x,i})$ is the item text at position $i$ in $x$.
We update the evidence log by merging the newly observed entries:
\begin{equation}
\mathcal{E}(x)\leftarrow \mathrm{Merge}\big(\mathcal{E}(x), \{e^{(r)}_{x,i}\}_{i\in\mathcal{H}(x)}\big),
\label{eq:evidence_update}
\end{equation}
where $\mathrm{Merge}(\cdot)$ de-duplicates repeated positions and keeps the most recent $L$ traces per position. As a result, $\mathcal{E}(x)$ provides stable anchors together with quantitative disagreement trajectories, so the LLM can concentrate on consistently contentious positions.

With $\mathcal{E}(x)$ as a compact, position-aligned summary, we query the LLM to update semantic priors on the hard cases of each selected sequence.
For each $x\in\mathcal{P}$, we query the LLM for updated semantic scores on hard cases:
\begin{equation}
\{p^{\,\text{new}}_{\mathrm{sem}}[i]\}_{i\in\mathcal{H}(x)}
=\mathcal{M}_\phi\Big(\{\mathcal{T}(s_{x,j})\}_{j=1}^{n},\ \mathcal{E}(x)\Big).
\label{eq:llm_refresh}
\end{equation}
The query is formatted as a structured record (see Appendix~\ref{app:calibration}) by injecting the evidence log $\mathcal{E}(x)$ into the prompt, while keeping the same scoring protocol as in Sec.~\ref{sec:method_semantic_prior}. We then write back the refreshed priors by
\begin{equation}
p_{\mathrm{sem}}[i]\leftarrow p^{\,\text{new}}_{\mathrm{sem}}[i],\quad \forall i\in\mathcal{H}(x),\ x\in\mathcal{P},
\label{eq:writeback}
\end{equation}
while keeping other positions unchanged.

\subsection{Training Objective}
\label{subsec:training_objective}

We optimize the backbone for next-item prediction with $\mathcal{L}_{\mathrm{rec}}$, and add two auxiliary regularizers:
\begin{equation}
\mathcal{L}
=
\mathcal{L}_{\mathrm{rec}}
+
\lambda_{\mathrm{aux}}
\Big(
\mathcal{L}_{\mathrm{risk\text{-}attn}}
+
\lambda_{\mathrm{cons}}\,
\mathcal{L}_{\mathrm{cons}}
\Big).
\label{eq:overall_obj}
\end{equation}
$\mathcal{L}_{\mathrm{risk\text{-}attn}}$ discourages allocating attention mass to high-risk positions.
In addition, we introduce a masked consistency term to calibrate the model-side risk estimator on positions where the two views already provide consistent evidence, while leaving hard cases untouched for disagreement-guided refresh.

For a training sequence $x$ with valid history positions $\mathcal{V}(x)$, we define the mask
\begin{equation}
m_x[i]=\mathbb{I}\!\left[d_x[i]\le \delta\right],\qquad i\in\mathcal{V}(x),
\label{eq:mask}
\end{equation}
and apply a masked binary cross-entropy:
\begin{equation}
\mathcal{L}_{\mathrm{cons}}
=
\frac{\sum_{i\in\mathcal{V}(x)} m_x[i]\,
\mathrm{BCE}\!\left(p_{\mathrm{mod},x}[i],\,p_{\mathrm{sem},x}[i]\right)}
{\sum_{i\in\mathcal{V}(x)} m_x[i]+\epsilon}.
\label{eq:consistency}
\end{equation}
Between refresh rounds, $p_{\mathrm{sem}}$ is treated as fixed evidence, so $\mathcal{L}_{\mathrm{cons}}$ only calibrates the model-side risk output $p_{\mathrm{mod}}$.

\begin{table*}[t]
\centering
\caption{Overall performance on three Amazon domains under \textbf{10\% Noise} and \textbf{Original} settings. We report NDCG@K and HR@K with $K\in\{5,10,20\}$. Best results are in bold, and second-best are underlined.}
\label{tab:overall}
\begin{adjustbox}{width=\textwidth}
\footnotesize
\begin{tabular}{l l
    >{\small\selectfont}c >{\small\selectfont}c >{\small\selectfont}c
    >{\small\selectfont}c >{\small\selectfont}c >{\small\selectfont}c
    >{\small\selectfont}c >{\small\selectfont}c >{\small\selectfont}c
    >{\small\selectfont}c >{\small\selectfont}c >{\small\selectfont}c}
\toprule
\textbf{Dataset} & \textbf{Method}
& \multicolumn{6}{c}{\textbf{10\% Noise}}
& \multicolumn{6}{c}{\textbf{Original}} \\
\cmidrule(lr){3-8}\cmidrule(lr){9-14}
& 
& \textbf{NDCG@5} & \textbf{NDCG@10} & \textbf{NDCG@20} & \textbf{HR@5} & \textbf{HR@10} & \textbf{HR@20}
& \textbf{NDCG@5} & \textbf{NDCG@10} & \textbf{NDCG@20} & \textbf{HR@5} & \textbf{HR@10} & \textbf{HR@20} \\
\midrule

\multirow{7}{*}{Movie}
& SASRec   & 0.0295 & 0.0335 & 0.0378 & \underline{0.0422} & 0.0550 & 0.0722 & \underline{0.0365} & \underline{0.0413} & \underline{0.0458} & \underline{0.0534} & \underline{0.0680} & 0.0860 \\
& BERT4Rec & 0.0188 & 0.0230 & 0.0283 & 0.0270 & 0.0402 & 0.0612 & 0.0285 & 0.0327 & 0.0378 & 0.0386 & 0.0516 & 0.0718 \\
& CL4Rec   & 0.0295 & \underline{0.0340} & \underline{0.0388} & 0.0390 & 0.0532 & 0.0722 & 0.0314 & 0.0366 & 0.0423 & 0.0418 & 0.0580 & 0.0808 \\
& FMLP-Rec & 0.0294 & 0.0323 & 0.0357 & 0.0418 & 0.0510 & 0.0648 & 0.0356 & 0.0389 & 0.0435 & 0.0508 & 0.0610 & 0.0792 \\
& LLM4DSR  & \underline{0.0296} & 0.0339 & 0.0380 & 0.0420 & \underline{0.0552} & 0.0716 & 0.0344 & 0.0398 & 0.0455 & 0.0482 & 0.0650 & \textbf{0.0876} \\
& IADSR    & 0.0286 & 0.0331 & 0.0383 & 0.0406 & 0.0548 & \textbf{0.0758} & 0.0346 & 0.0403 & 0.0453 & 0.0488 & 0.0670 & \underline{0.0870} \\
& DC4SR  & \textbf{0.0300} & \textbf{0.0345} & \textbf{0.0390} & \textbf{0.0430} & \textbf{0.0568} & \underline{0.0752} & \textbf{0.0372} & \textbf{0.0420} & \textbf{0.0466} & \textbf{0.0544} & \textbf{0.0692} & \textbf{0.0876} \\
\midrule

\multirow{7}{*}{Office}
& SASRec   & 0.0070 & \underline{0.0114} & 0.0148 & 0.0135 & \underline{0.0270} & 0.0405 & \underline{0.0159} & \underline{0.0206} & \underline{0.0259} & \underline{0.0286} & \underline{0.0431} & 0.0644 \\
& BERT4Rec & 0.0030 & 0.0038 & 0.0058 & 0.0047 & 0.0073 & 0.0151 & 0.0051 & 0.0065 & 0.0081 & 0.0094 & 0.0135 & 0.0203 \\
& CL4Rec   & \underline{0.0082} & 0.0112 & \underline{0.0155} & \underline{0.0161} & 0.0255 & 0.0426 & 0.0041 & 0.0072 & 0.0119 & 0.0068 & 0.0166 & 0.0358 \\
& FMLP-Rec & 0.0055 & 0.0101 & 0.0152 & 0.0119 & 0.0265 & \underline{0.0468} & 0.0092 & 0.0131 & 0.0193 & 0.0182 & 0.0301 & 0.0545 \\
& LLM4DSR  & 0.0045 & 0.0065 & 0.0114 & 0.0088 & 0.0151 & 0.0343 & 0.0113 & 0.0187 & 0.0245 & 0.0197 & 0.0421 & \underline{0.0660} \\
& IADSR    & 0.0046 & 0.0088 & 0.0147 & 0.0099 & 0.0229 & 0.0462 & 0.0080 & 0.0136 & 0.0210 & 0.0161 & 0.0338 & 0.0634 \\
& DC4SR  & \textbf{0.0107} & \textbf{0.0140} & \textbf{0.0185} & \textbf{0.0208} & \textbf{0.0317} & \textbf{0.0488} & \textbf{0.0161} & \textbf{0.0212} & \textbf{0.0263} & \textbf{0.0291} & \textbf{0.0447} & \textbf{0.0691} \\
\midrule

\multirow{7}{*}{Toys}
& SASRec   & 0.0082 & 0.0113 & 0.0146 & 0.0163 & 0.0257 & 0.0388 & \underline{0.0106} & \underline{0.0147} & \underline{0.0191} & \underline{0.0215} & \underline{0.0340} & \underline{0.0513} \\
& BERT4Rec & 0.0057 & 0.0076 & 0.0097 & 0.0092 & 0.0152 & 0.0234 & 0.0059 & 0.0072 & 0.0092 & 0.0081 & 0.0119 & 0.0198 \\
& CL4Rec   & 0.0070 & 0.0094 & 0.0130 & 0.0138 & 0.0213 & 0.0357 & 0.0086 & 0.0112 & 0.0155 & 0.0152 & 0.0236 & 0.0405 \\
& FMLP-Rec & \underline{0.0087} & 0.0112 & 0.0127 & 0.0167 & 0.0242 & 0.0300 & 0.0053 & 0.0064 & 0.0081 & 0.0102 & 0.0138 & 0.0202 \\
& LLM4DSR  & 0.0072 & 0.0102 & 0.0133 & 0.0140 & 0.0236 & 0.0359 & 0.0087 & 0.0129 & 0.0165 & 0.0169 & 0.0298 & 0.0442 \\
& IADSR    & \underline{0.0087} & \underline{0.0121} & \underline{0.0151} & \underline{0.0169} & \underline{0.0275} & \underline{0.0394} & 0.0076 & 0.0111 & 0.0148 & 0.0144 & 0.0252 & 0.0403 \\
& DC4SR  & \textbf{0.0092} & \textbf{0.0125} & \textbf{0.0156} & \textbf{0.0179} & \textbf{0.0286} & \textbf{0.0407} & \textbf{0.0120} & \textbf{0.0162} & \textbf{0.0202} & \textbf{0.0240} & \textbf{0.0373} & \textbf{0.0530} \\
\bottomrule
\end{tabular}
\end{adjustbox}
\end{table*}

\section{Experiments}
We design our experiments to answer the following research questions (RQs):
\noindent\textbf{RQ1:} Does DC4SR improve recommendation quality under both synthetic noise and clean interaction logs?
\noindent\textbf{RQ2:} Does semantic--sequential disagreement persist across training checkpoints, and can it provide actionable signals for denoising?
\noindent\textbf{RQ3:} Which components and design choices of DC4SR contribute most to the performance gains?
\noindent\textbf{RQ4:} How sensitive is DC4SR to design choices, including the semantic-side LLM backbone and key hyperparameters (e.g., $\alpha$, $(\tau_{\mathrm{low}},\tau_{\mathrm{high}})$, and the refresh interval $R$)?

\subsection{Experimental Settings}
\noindent\textbf{Datasets.}
We conduct experiments on three Amazon domains~\cite{he2016ups}: Movie, Toys and Games, and Office.
User interactions are organized as chronological sequences, and item titles are used as the textual input for semantic denoising.
Additional dataset statistics are reported in Appendix~\ref{app:dataset}.\par
\noindent\textbf{Baselines.} We compare DC4SR with three groups of baselines.
(i) Traditional sequential recommendation: SASRec~\cite{kang2018self} is a self-attentive Transformer recommender, and BERT4Rec~\cite{sun2019bert4rec} uses a bidirectional Transformer trained with masked item prediction.
(ii) Traditional denoising methods: CL4Rec~\cite{xie2022contrastive} enhances robustness via contrastive denoising with sequence augmentations, while FMLP-Rec~\cite{zhou2022filter} adopts a filter-enhanced MLP architecture that serves as a strong and efficient backbone with implicit regularization on noisy patterns.
(iii) LLM-assisted denoising: LLM4DSR~\cite{wang2025llm4dsr} uses LLM-derived semantic judgments to detect and cleanse suspicious interactions, and IADSR~\cite{wu2025empowering} exploits LLM embeddings/semantics as denoising signals to suppress noisy interactions during training.
\par

\noindent\textbf{Evaluation protocol.} We follow the next-item prediction setting implemented in our code.
For each test sequence, we compute scores over the full item set using the model prediction head, rank all items by sorting,
and obtain the rank of the ground-truth next item.
We report HR@K and NDCG@K for $K\in\{5,10,20\}$.
NDCG is computed as $1/\log_2(\mathrm{rank}+2)$ (with 0-indexed rank) when the target is within top-$K$, and zero otherwise.\par

\noindent\textbf{Implementation details.} We train with Adam and a point-wise BCE loss; positives are obtained by shifting each sequence by one position, and negatives are sampled from the full item set while excluding sequence items and the padding index. During training, we feed position-wise noise signals for historical positions, applied only to the history segment. The model outputs positive and negative logits, optionally with an auxiliary denoising loss added to the BCE loss. The LLM uses Llama-3 3B \cite{touvron2023llama} with a LoRA adapter \cite{hu2022lora} fine-tuned for denoising instructions. 
For the 10\% noisy setting, we corrupt training histories by randomly replacing each interaction with probability 0.1 while keeping test targets unchanged.
Every $R$ epochs, we refresh semantic priors on the top-$M$ most unreliable sequences (up to $K$ positions each) and continue training while preserving the optimizer state.
We fix $(M,K,R)=(200,3,10)$ for all datasets; the calibration adds modest overhead, does not affect evaluation, and the motivation probe is used for analysis only.

\subsection{Overall Performance}
Table~\ref{tab:overall} reports the results of DC4SR and all baselines on three Amazon domains under clean and 10\% noisy settings.
Across datasets, DC4SR achieves the best or second-best performance, indicating that combining an LLM-based semantic prior with model-side training signals improves robustness for Transformer-based sequential recommendation.
Compared with conventional recommenders (SASRec, BERT4Rec) and traditional denoising methods (CL4Rec, FMLP-Rec), DC4SR yields more consistent improvements across datasets under both settings.
Compared with LLM-assisted denoising baselines (LLM4DSR, IADSR), DC4SR further improves performance, consistent with the benefit of the calibration that prioritizes disputed positions for LLM review and prior refresh rather than relying on static semantic judgments.
Overall, these results support that disagreement-guided semantic denoising leads to more reliable recommendation quality in the presence of noisy interactions.

\begin{figure}[t]
  \centering
  \includegraphics[width=0.96\linewidth]{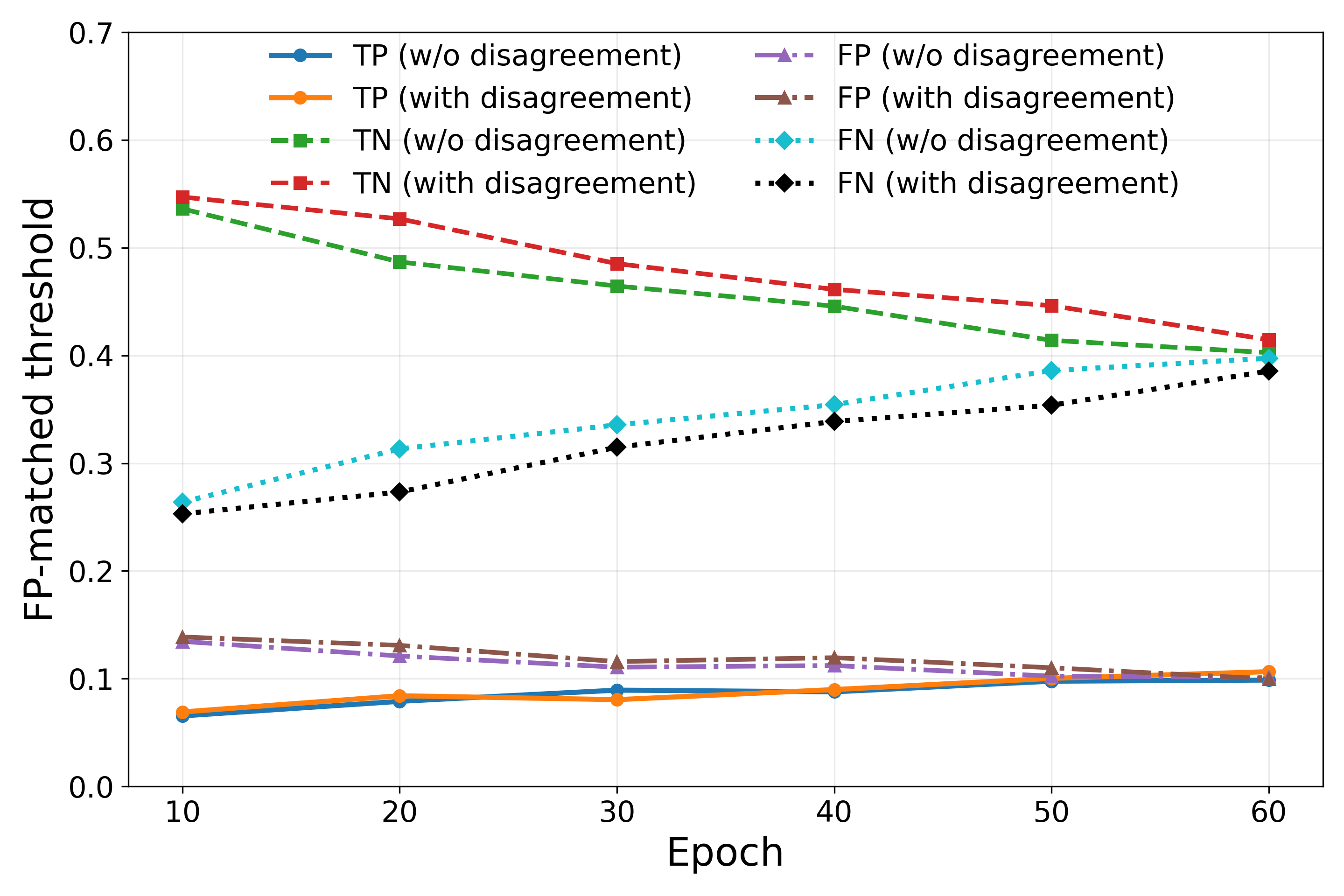}
  \caption{Four outcome fractions (TP/FP/FN/TN) when using the LLM semantic score $S_t$ to flag suspicious interactions on Office.}
  \label{fig:semantic_quadrants}
\end{figure}

\begin{table}[t]
\centering
\caption{Ablation results on three Amazon domains. We report NDCG@K and HR@K with $K \in \{5,10,20\}$ on the 10\% noise setting. Best results are in bold and second-best are underlined.}
\label{tab:ablation}

\setlength{\tabcolsep}{2.0pt}
\renewcommand{\arraystretch}{1.00}
\small

\begin{adjustbox}{width=\columnwidth}
\begin{tabular}{l l c c c c c c}
\toprule
\multirow{2}{*}{\textbf{Variant}} & \multirow{2}{*}{\textbf{Variant}}
& \multicolumn{3}{c}{\textbf{NDCG}} & \multicolumn{3}{c}{\textbf{HR}} \\
\cmidrule(lr){3-5}\cmidrule(lr){6-8}
& & \textbf{@5} & \textbf{@10} & \textbf{@20} & \textbf{@5} & \textbf{@10} & \textbf{@20} \\
\midrule
\multirow{7}{*}{Toys}
& \textsc{Ours} (Full)    & \textbf{0.0092} & \textbf{0.0125} & \textbf{0.0156} & \textbf{0.0179} & \textbf{0.0286} & \textbf{0.0407} \\
& w/o Calibration         & 0.0079 & 0.0107 & 0.0140 & 0.0154 & 0.0242 & 0.0371 \\
& w/o LLM                 & \underline{0.0082} & \underline{0.0115} & \underline{0.0145} & 0.0156 & \underline{0.0259} & \underline{0.0378} \\
& w/o Posterior           & 0.0072 & 0.0091 & 0.0120 & 0.0134 & 0.0192 & 0.0311 \\
& w/o Risk-Attn           & 0.0079 & 0.0107 & 0.0139 & \underline{0.0159} & 0.0244 & 0.0371 \\
& w/o Consistency         & 0.0068 & 0.0083 & 0.0111 & 0.0127 & 0.0175 & 0.0288 \\
& uniform-$w_h$           & 0.0052 & 0.0075 & 0.0099 & 0.0100 & 0.0171 & 0.0265 \\
\midrule
\multirow{7}{*}{Office}
& \textsc{Ours} (Full)    & \textbf{0.0107} & \textbf{0.0140} & \textbf{0.0185} & \textbf{0.0208} & \textbf{0.0317} & \textbf{0.0488} \\
& w/o Calibration         & \underline{0.0099} & \underline{0.0136} & \underline{0.0170} & \underline{0.0197} & \underline{0.0312} & \underline{0.0452} \\
& w/o LLM                 & 0.0088 & 0.0116 & 0.0159 & 0.0161 & 0.0249 & 0.0421 \\
& w/o Posterior           & 0.0036 & 0.0069 & 0.0104 & 0.0068 & 0.0171 & 0.0312 \\
& w/o Risk-Attn           & 0.0073 & 0.0109 & 0.0144 & 0.0125 & 0.0239 & 0.0379 \\
& w/o Consistency         & 0.0030 & 0.0037 & 0.0065 & 0.0057 & 0.0078 & 0.0192 \\
& uniform-$w_h$           & 0.0015 & 0.0031 & 0.0078 & 0.0031 & 0.0078 & 0.0265 \\
\midrule
\multirow{7}{*}{Movie}
& \textsc{Ours} (Full)    & \textbf{0.0300} & \textbf{0.0345} & \textbf{0.0390} & \textbf{0.0430} & \textbf{0.0568} & \textbf{0.0752} \\
& w/o Calibration         & \underline{0.0299} & 0.0338 & 0.0379 & \underline{0.0426} & 0.0550 & 0.0712 \\
& w/o LLM                 & 0.0298 & \underline{0.0340} & \underline{0.0389} & \underline{0.0426} & \underline{0.0556} & \underline{0.0748} \\
& w/o Posterior           & 0.0276 & 0.0324 & 0.0370 & 0.0398 & 0.0546 & 0.0732 \\
& w/o Risk-Attn           & 0.0295 & \underline{0.0340} & 0.0383 & 0.0422 & \underline{0.0562} & 0.0732 \\
& w/o Consistency         & 0.0285 & 0.0323 & 0.0367 & 0.0412 & 0.0532 & 0.0710 \\
& uniform-$w_h$           & 0.0279 & 0.0328 & 0.0372 & 0.0400 & 0.0550 & 0.0730 \\
\bottomrule
\end{tabular}
\end{adjustbox}
\end{table}

\subsection{Disagreement Analysis}
\label{sec:case_study_semantic_drift}
We conduct a dataset-level diagnostic on Office to quantify the mismatch between LLM-based semantic denoising and model-side training risk.
We sample interaction positions $t$ from validation sequences and compute an LLM semantic score $S_t$ for interaction $i_t$ under the current backbone state.
We treat $S_t\ge\theta_e$ as predicting harmful and define TP/FP/FN/TN by comparing this prediction with a harm label, and harmful represents the positive class.
The harm label is defined by a counterfactual masking test:
\begin{equation}
  R_t = M(x\setminus i_t) - M(x),
\end{equation}
where $x$ is the original sequence and $x\setminus i_t$ masks the interaction at position $t$, and $M(\cdot)$ is a ranking metric NDCG@K computed on the same validation samples.
If masking $i_t$ improves the metric, then $i_t$ hurts ranking performance under the current backbone and is therefore harmful; we treat $R_t>0$ as harmful.

To keep a comparable operating point across training stages, we choose $\theta_e$ to control a fixed false-positive rate.
At an anchor epoch $e_{\mathrm{anc}}$, we set $\theta_{\mathrm{anc}}$ and define
\begin{equation}
  \alpha=\Pr\!\left(S_t\ge\theta_{\mathrm{anc}}\wedge R_t\le 0\right).
\end{equation}
For each epoch $e$, we choose $\theta_e$ such that
\begin{equation}
  \Pr\!\left(S_t\ge\theta_e\wedge R_t\le 0\right)\le \alpha.
\end{equation}
Figure~\ref{fig:semantic_quadrants} shows a persistent FP fraction, indicating that semantic-only filtering would over-remove interactions that are not harmful under the backbone's current optimization state.
Meanwhile, the FN fraction remains non-trivial, indicating that some training-time harmful interactions are not captured by semantics alone even under a matched operating point.

Together, these observations reveal a persistent semantic and sequential mismatch under the learned backbone.
This motivates leveraging semantic and sequential disagreement for denoising. We therefore treat $S_t$ as an external semantic prior and $R_t$ as an optimization-state-dependent risk signal, and use their disagreement as a complementary cue for denoising.

\subsection{Ablation Study}
\label{sec:ablation}
We conduct ablations to quantify the contribution of key components in \textsc{Ours}.
The full model combines an LLM semantic prior $p_{\mathrm{sem}}$, a model-side posterior $p_{\mathrm{mod}}$, a disagreement-driven prior refresh, and a risk-aware attention regularizer with traction-informed head weights.
Table~\ref{tab:ablation} reports the following variants: w/o Calibration disables the refresh and write-back step; w/o LLM removes $p_{\mathrm{sem}}$; w/o Posterior removes $p_{\mathrm{mod}}$; w/o Risk-Attn removes the regularizer; w/o Consistency removes $\mathcal{L}_{\mathrm{cons}}$ only; 
and uniform-$w_h$ disables head-specific weighting by setting $w_h{=}0$ for all heads, yielding an unweighted risk-aware attention regularizer.

We report NDCG@K and HR@K on three Amazon domains with $K\in\{5,10,20\}$.
Removing the posterior causes the largest degradation, and removing the semantic pathway also hurts on Office and Toys, indicating that the two-view signals are complementary.
Disabling prior refresh or consistency alignment reduces performance across domains, and removing risk-aware attention further degrades results; using uniform head weights amplifies this drop.

We further examine whether DC4SR is sensitive to the fusion rule used to construct the training-time weight $c[i]\in(0,1)$ from $p_{\mathrm{mod}}[i]$, $p_{\mathrm{sem}}[i]$, and their disagreement $d[i]$.
Keeping the gating function $s(\cdot)$ (defined by $\tau_{\mathrm{low}}$ and $\tau_{\mathrm{high}}$) and all other settings unchanged, we replace the default clipped linear interpolation in Eq.~(12) with three alternatives: Product-of-Experts (PoE), a lightweight learnable fusion network (FusNet), and a Bayesian-style odds update.
Table~\ref{tab:fusion_compare} suggests that the clipped linear fusion is the most stable option, delivering best or second-best results across datasets and cutoffs in most cases.
Alternative operators show sporadic gains on specific metrics but lack consistent improvements. 

\begin{figure}[t]
  \centering
  \setlength{\tabcolsep}{2pt}
  \begin{tabular}{cc}
    \subcaptionbox{risk-fusion coefficient $\alpha$.\label{fig:sens_alpha}}{
      \includegraphics[width=0.48\linewidth]{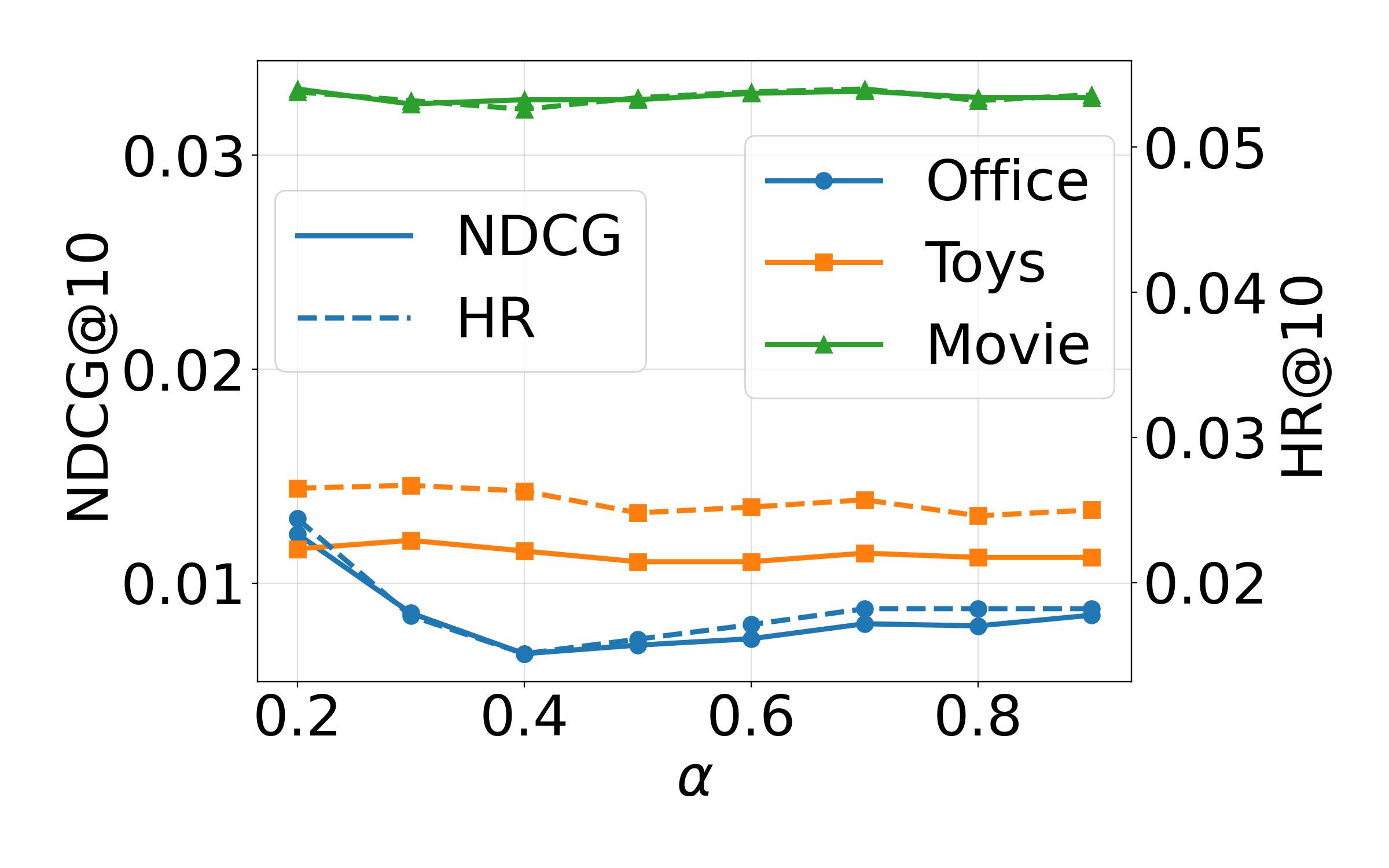}
    } &
    \subcaptionbox{refresh interval $R$.\label{fig:sens_R}}{
      \includegraphics[width=0.48\linewidth]{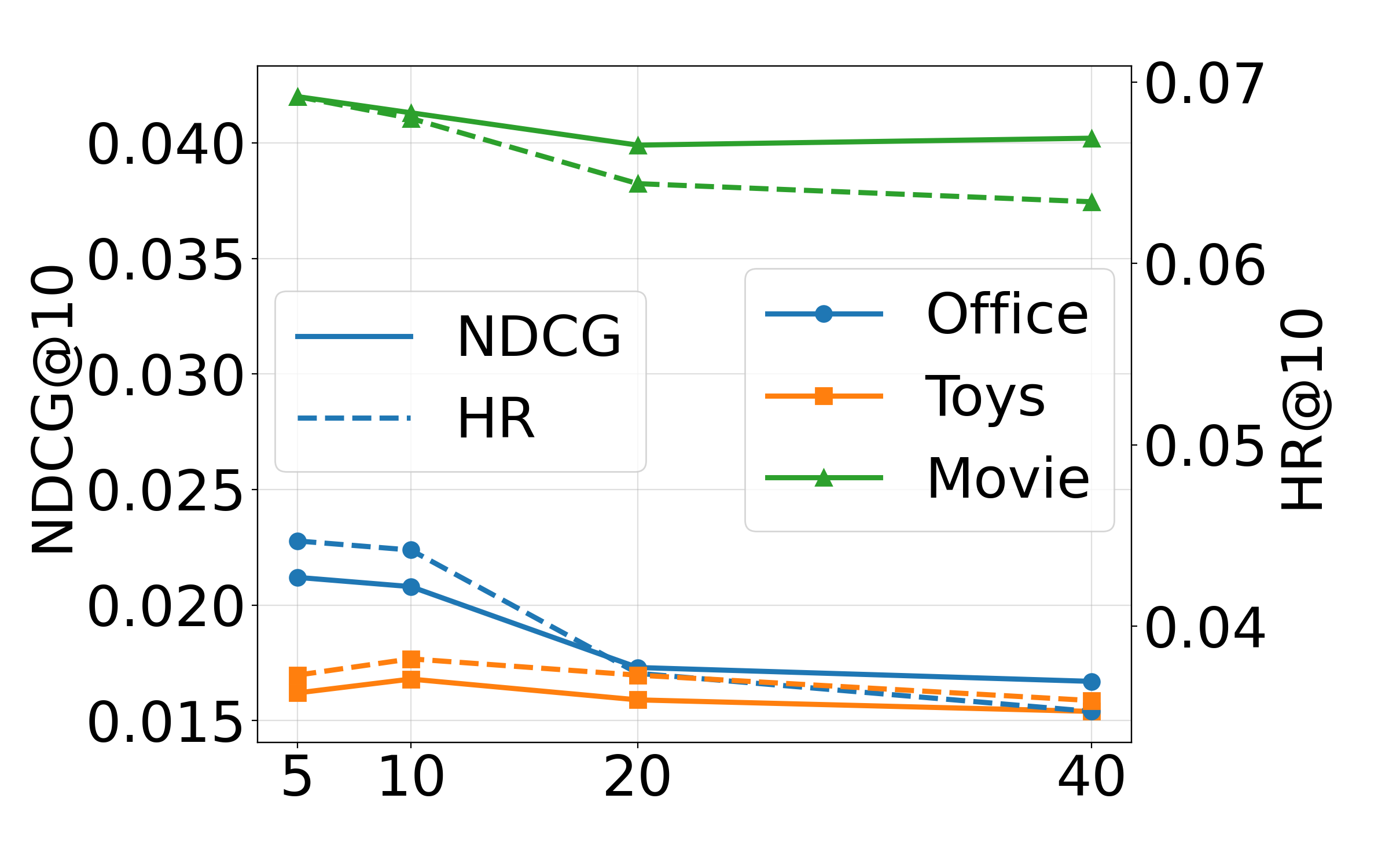}
    } \\

    \subcaptionbox{$\tau_{\mathrm{high}}$ with $\tau_{\mathrm{low}}{=}0.2$.\label{fig:sens_tauhigh}}{
      \includegraphics[width=0.48\linewidth]{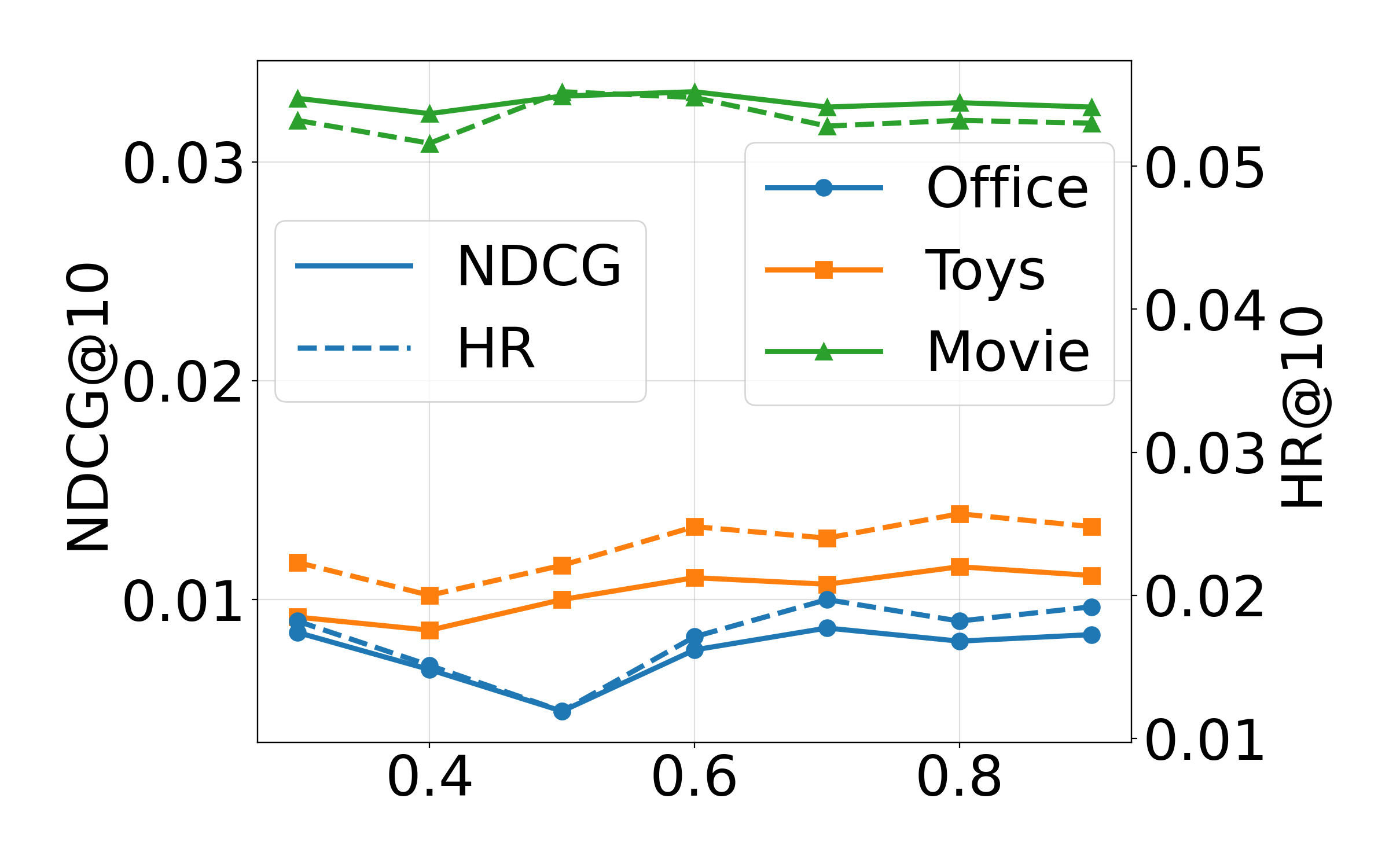}
    } &
    \subcaptionbox{$\tau_{\mathrm{low}}$ with $\tau_{\mathrm{high}}{=}0.7$.\label{fig:sens_taulow}}{
      \includegraphics[width=0.48\linewidth]{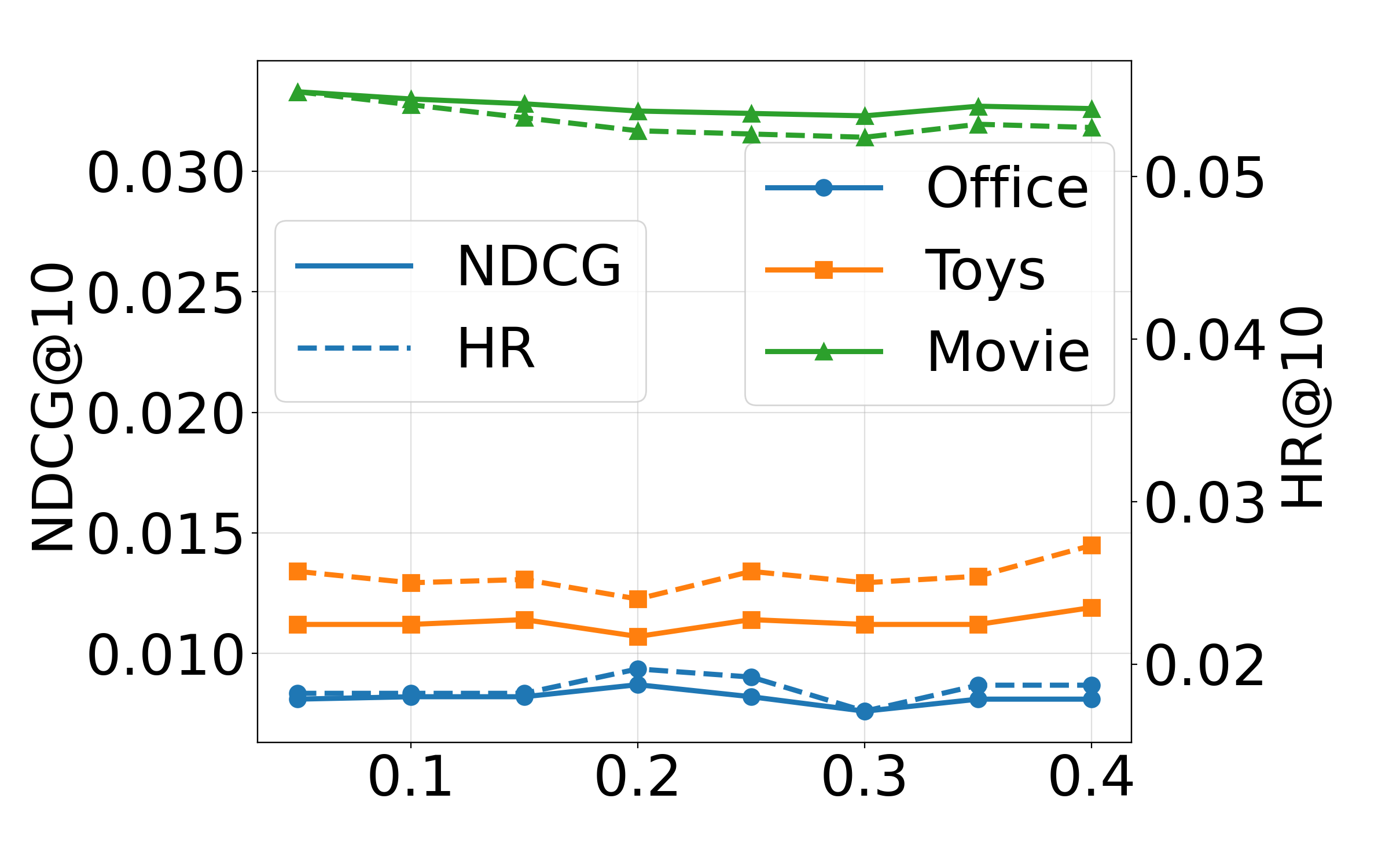}
    } \\
  \end{tabular}
  \caption{Sensitivity to $\alpha$, $R$, and $(\tau_{\mathrm{low}},\tau_{\mathrm{high}})$ on Movie, Toys, and Office. Solid lines denote NDCG@10 and dashed lines denote HR@10.}
  \label{fig:sensitivity_all}
\end{figure}

\begin{table}[t]
\centering
\caption{Comparison of fusion operators for constructing the training-time weight $c[i]$ on the original Office, Movie, and Toy datasets. Best results are in bold, and second-best are underlined.}
\footnotesize
\setlength{\tabcolsep}{3.0pt}
\renewcommand{\arraystretch}{1.05}
\begin{adjustbox}{width=\columnwidth}
\begin{tabular}{l l c c c c c c}
\toprule
\multirow{2}{*}{\textbf{Dataset}} & \multirow{2}{*}{\textbf{Fusion}}
& \multicolumn{3}{c}{\textbf{NDCG}} & \multicolumn{3}{c}{\textbf{HR}} \\
\cmidrule(lr){3-5}\cmidrule(lr){6-8}
& & \textbf{@5} & \textbf{@10} & \textbf{@20} & \textbf{@5} & \textbf{@10} & \textbf{@20} \\
\midrule
\multirow{4}{*}{Office}
& DC4SR    & \textbf{0.0161} & \textbf{0.0212} & \underline{0.0263} & \textbf{0.0291} & \textbf{0.0447} & 0.0691 \\
& PoE       & 0.0130 & 0.0177 & 0.0242 & 0.0234 & 0.0384 & 0.0644 \\
& FusNet    & 0.0131 & 0.0196 & \textbf{0.0276} & 0.0229 & 0.0426 & \textbf{0.0748} \\
& BayesOdds & 0.0119 & 0.0182 & 0.0260 & 0.0203 & 0.0400 & \underline{0.0712} \\
\midrule
\multirow{4}{*}{Movie}
& DC4SR    & \textbf{0.0372} & \textbf{0.0420} & \textbf{0.0466} & \textbf{0.0544} & \textbf{0.0692} & \underline{0.0876} \\
& PoE       & 0.0355 & 0.0402 & 0.0453 & 0.0512 & \underline{0.0658} & 0.0860 \\
& FusNet    & 0.0356 & 0.0397 & \underline{0.0459} & 0.0510 & 0.0638 & \textbf{0.0884} \\
& BayesOdds & 0.0354 & \underline{0.0404} & 0.0453 & 0.0506 & \underline{0.0658} & 0.0854 \\
\midrule
\multirow{4}{*}{Toy}
& DC4SR    & \textbf{0.0120} & \textbf{0.0162} & \textbf{0.0202} & \textbf{0.0240} & \textbf{0.0373} & \textbf{0.0530} \\
& PoE       & 0.0103 & 0.0139 & 0.0182 & 0.0202 & 0.0313 & 0.0482 \\
& FusNet    & \underline{0.0110} & \underline{0.0153} & \underline{0.0188} & \underline{0.0219} & \underline{0.0353} & \underline{0.0494} \\
& BayesOdds & 0.0093 & 0.0129 & 0.0164 & 0.0179 & 0.0290 & 0.0432 \\
\bottomrule
\end{tabular}
\end{adjustbox}
\label{tab:fusion_compare}
\end{table}

\subsection{LLM Backbone Analysis}
\label{sec:exp_llm_backbone}

To examine the sensitivity of DC4SR to the semantic-side LLM, we instantiate DC4SR with two 3B-scale backbones, Llama-3B and Qwen-3B \cite{bai2023qwen}, while keeping all other training settings unchanged.
We evaluate under the original setting and compare against two LLM-assisted denoising baselines, LLM4DSR and IADSR, both using Qwen-3B for semantic guidance.
Table~\ref{tab:llm_backbone_clean} reports the results.

Overall, DC4SR achieves comparable performance across the two 3B-scale LLMs, indicating that the framework is not overly sensitive to the specific LLM backbone and that the semantic signal provides a stable prior.
Under the same semantic-side LLM setting, DC4SR consistently outperforms LLM-assisted baselines that rely on static semantic signals without disagreement-guided calibration, highlighting the benefit of combining semantic judgments with model-side signals to prioritize intervention-worthy positions.
Due to limited computational resources, we only compare 3B-scale LLMs in this study; evaluating larger LLMs will help further assess both the robustness of the framework to backbone choices and its potential scalability with stronger semantic models (e.g., different LoRA setups, quantization, and prompt formats).

\subsection{Hyperparameter Sensitivity}
\label{sec:sensitivity}
Figure~\ref{fig:sensitivity_all} reports the sensitivity of DC4SR to $\alpha$, $(\tau_{\mathrm{low}},\tau_{\mathrm{high}})$, and the refresh interval $R$ on Movie, Toys, and Office.
Overall, Movie and Toys remain comparatively stable across the tested ranges, whereas Office is more sensitive, showing a consistent preference for smaller $\alpha$ and more noticeable metric fluctuations when varying $\tau_{\mathrm{high}}$ or $\tau_{\mathrm{low}}$.
For $R$, we observe refresh intervals yield stable gains, while overly frequent refreshes may introduce additional variance due to repeated prior updates, and overly sparse refreshes may delay the correction of hard, noisy positions.
This cross-domain heterogeneity is plausible given differences in interaction sparsity and how quickly semantic priors need to track model-side signals.

For a unified and comparable evaluation, we fix $\alpha=0.8$, $\tau_{\mathrm{low}}=0.2$, $\tau_{\mathrm{high}}=0.7$, and $R=10$ in all experiments; since the optimal hyperparameters can be dataset-specific, we prioritize consistent reporting over per-domain tuning.

\begin{table}[t]
\centering
\caption{Performance under the original setting with different LLM backbones. We report NDCG@K and HR@K with $K\in\{5,10,20\}$. Best results are in bold, and second-best are underlined.}
\label{tab:llm_backbone_clean}

\setlength{\tabcolsep}{2.0pt}
\renewcommand{\arraystretch}{1.00}
\small

\begin{adjustbox}{width=\columnwidth}
\begin{tabular}{l l c c c c c c}
\toprule
\multirow{2}{*}{\textbf{Dataset}} & \multirow{2}{*}{\textbf{Method}}
& \multicolumn{3}{c}{\textbf{NDCG}} & \multicolumn{3}{c}{\textbf{HR}} \\
\cmidrule(lr){3-5}\cmidrule(lr){6-8}
& & \textbf{@5} & \textbf{@10} & \textbf{@20} & \textbf{@5} & \textbf{@10} & \textbf{@20} \\
\midrule
\multirow{4}{*}{Movie}
& DC4SR (Llama-3B) & \textbf{0.0372} & \textbf{0.0420} & \textbf{0.0466} & \textbf{0.0544} & \textbf{0.0692} & \textbf{0.0876} \\
& DC4SR (Qwen-3B)  & \underline{0.0365} & \underline{0.0413} & \underline{0.0464} & \underline{0.0534} & \underline{0.0680} & \underline{0.0860} \\
& LLM4DSR (Qwen-3B)  & 0.0355 & 0.0402 & 0.0458 & 0.0490 & 0.0634 & 0.0850 \\
& IADSR (Qwen-3B)    & 0.0353 & 0.0399 & 0.0445 & 0.0502 & 0.0644 & 0.0826 \\
\midrule

\multirow{4}{*}{Toys}
& DC4SR (Llama-3B) & \underline{0.0120} & \underline{0.0162} & \textbf{0.0202} & \underline{0.0240} & \underline{0.0373} & \textbf{0.0530} \\
& DC4SR (Qwen-3B)  & \textbf{0.0123} & \textbf{0.0168} & \textbf{0.0202} & \textbf{0.0244} & \textbf{0.0382} & \underline{0.0521} \\
& LLM4DSR (Qwen-3B)  & 0.0091 & 0.0133 & 0.0171 & 0.0165 & 0.0296 & 0.0446 \\
& IADSR (Qwen-3B)    & 0.0104 & 0.0154 & \underline{0.0191} & 0.0207 & 0.0359 & 0.0505 \\
\midrule

\multirow{4}{*}{Office}
& DC4SR (Llama-3B) & \textbf{0.0161} & \textbf{0.0212} & \underline{0.0263} & \textbf{0.0291} & \textbf{0.0447} & \textbf{0.0691} \\
& DC4SR (Qwen-3B)  & \underline{0.0150} & \underline{0.0208} & \textbf{0.0271} & \underline{0.0260} & \underline{0.0442} & \underline{0.0691} \\
& LLM4DSR (Qwen-3B)  & 0.0132 & 0.0173 & 0.0246 & 0.0249 & 0.0374 & 0.0665 \\
& IADSR (Qwen-3B)    & 0.0128 & 0.0174 & 0.0246 & 0.0239 & 0.0384 & 0.0675 \\
\bottomrule
\end{tabular}
\end{adjustbox}
\end{table}

\section{Related Work}
Sequential recommendation \cite{vaswani2017attention,kang2018self,sun2019bert4rec} is typically trained on large-scale implicit feedback logs, which inevitably contain noisy interactions \cite{joachims2007evaluating, zhang2025robust}, such as accidental clicks, short-term interest drifts, and externally induced jumps \cite{wang2023efficient, chen2023bias}. 
To mitigate this behavioral noise, a substantial line of work focuses on identifying suspicious interactions and suppressing their influence during training \cite{sun2021does}. 
Early approaches rely on filtering or reweighting signals, e.g., truncating or down-weighting high-loss interactions \cite{wang2021denoising}, learning hierarchical inconsistency to detect unreliable positions \cite{zhang2022hierarchical}, performing self-correction to repair sequences \cite{lin2023self}, or using contrastive objectives to separate clean patterns \cite{xie2022contrastive}. 
Subsequent studies emphasize that high loss is not equivalent to noise; it may also reflect hard-but-informative examples or optimization instability, and propose more stable dropping criteria and progressive correction mechanisms \cite{he2024double, zhang2024ssdrec}. 
Complementary directions leverage cross-model agreement \cite{wang2022learning} or contribution-aware pruning \cite{zhang2025shapley} to avoid discarding useful preference signals. 
Beyond data-level operations, several denoising sequential recommenders explicitly regularize noise amplification inside self-attention encoders~\cite{chen2022denoising,gao2022sgdl}, which is closely related to our head-level robustness perspective.

With the rise of LLMs~\cite{zhao2023survey, kasneci2023chatgpt, li2024large, song2024large}, recent LLM-based denoising methods exploit their strong semantic understanding and commonsense reasoning capabilities to derive semantic signals from item descriptions and interaction contexts~\cite{wang2025mllm4rec, zhang2023denoising, liu2025large, boz2025improving}. 
These methods aim to identify suspicious behaviors from a text-semantic perspective, providing more interpretable and semantically grounded denoising cues than purely heuristic or loss-based criteria~\cite{zhang2023denoising, wang2025llm4dsr, hu2024semanticembedding}. 
In practice, semantic signals are often applied as a pre-processing step via static sequence rewriting, interaction removal, or soft reweighting~\cite{wang2025llm4dsr, wu2025empowering, sun2025llm4rsr}. 
However, such semantic judgments rely on a global semantic view and remain decoupled from the backbone’s evolving optimization state, where the influence of a noisy interaction can change across training stages; consequently, semantic anomaly does not necessarily imply training-time harmfulness, motivating learning-aware denoising that incorporates model-side training dynamics~\cite{zhang2025shapley, ji2023survey}.

\section{Conclusion}
We propose DC4SR, a \underline{D}ual-view \underline{C}alibration framework that combines a dynamic LLM semantic prior with model-side training signals for \underline{S}equential \underline{R}ecommendation denoising.
Experiments on three Amazon domains under original and 10\% noisy settings show consistent gains over strong baselines, and disagreement analysis confirms complementary semantic and model signals across training stages.
DC4SR further mitigates traction heads via head-aware regularization and revisits ambiguous positions instead of static semantic edits.

\begin{acks}
To Robert, for the bagels and explaining CMYK and color spaces.
\end{acks}

\bibliographystyle{ACM-Reference-Format}
\bibliography{base}

@article{joachims2007evaluating,
  title={Evaluating the accuracy of implicit feedback from clicks and query reformulations in web search},
  author={Joachims, Thorsten and Granka, Laura and Pan, Bing and Hembrooke, Helene and Radlinski, Filip and Gay, Geri},
  journal={ACM Transactions on Information Systems (TOIS)},
  volume={25},
  number={2},
  pages={7--es},
  year={2007},
  publisher={ACM New York, NY, USA}
}

@inproceedings{wang2019sequential,
  title={Sequential Recommender Systems: Challenges, Progress and Prospects},
  author={Wang, Shoujin and Hu, Liang and Wang, Yan and Cao, Longbing and Sheng, Quan Z and Orgun, Mehmet},
  booktitle={Proceedings of the Twenty-Eighth International Joint Conference on Artificial Intelligence},
  pages={6332--6338},
  year={2019},
  organization={International Joint Conferences on Artificial Intelligence Organization}
}

@inproceedings{martinez2016managing,
  title={Managing natural noise in recommender systems},
  author={Mart{\'\i}nez, Luis and Castro, Jorge and Yera, Raciel},
  booktitle={International conference on theory and practice of natural computing},
  pages={3--17},
  year={2016},
  organization={Springer}
}

@inproceedings{wang2021clicks,
  title={Clicks can be cheating: Counterfactual recommendation for mitigating clickbait issue},
  author={Wang, Wenjie and Feng, Fuli and He, Xiangnan and Zhang, Hanwang and Chua, Tat-Seng},
  booktitle={Proceedings of the 44th international ACM SIGIR conference on research and development in information retrieval},
  pages={1288--1297},
  year={2021}
}

@inproceedings{wang2021denoising,
  title={Denoising implicit feedback for recommendation},
  author={Wang, Wenjie and Feng, Fuli and He, Xiangnan and Nie, Liqiang and Chua, Tat-Seng},
  booktitle={Proceedings of the 14th ACM international conference on web search and data mining},
  pages={373--381},
  year={2021}
}

@inproceedings{wang2022learning,
  title={Learning robust recommenders through cross-model agreement},
  author={Wang, Yu and Xin, Xin and Meng, Zaiqiao and Jose, Joemon M and Feng, Fuli and He, Xiangnan},
  booktitle={Proceedings of the ACM web conference 2022},
  pages={2015--2025},
  year={2022}
}

@inproceedings{kang2018self,
  title={Self-attentive sequential recommendation},
  author={Kang, Wang-Cheng and McAuley, Julian},
  booktitle={2018 IEEE international conference on data mining (ICDM)},
  pages={197--206},
  year={2018},
  organization={IEEE}
}

@inproceedings{sun2019bert4rec,
  title={BERT4Rec: Sequential recommendation with bidirectional encoder representations from transformer},
  author={Sun, Fei and Liu, Jun and Wu, Jian and Pei, Changhua and Lin, Xiao and Ou, Wenwu and Jiang, Peng},
  booktitle={Proceedings of the 28th ACM international conference on information and knowledge management},
  pages={1441--1450},
  year={2019}
}

@inproceedings{o2006detecting,
  title={Detecting noise in recommender system databases},
  author={O'Mahony, Michael P and Hurley, Neil J and Silvestre, Gu{\'e}nol{\'e} CM},
  booktitle={Proceedings of the 11th international conference on Intelligent user interfaces},
  pages={109--115},
  year={2006}
}

@inproceedings{lin2023self,
  title={A self-correcting sequential recommender},
  author={Lin, Yujie and Wang, Chenyang and Chen, Zhumin and Ren, Zhaochun and Xin, Xin and Yan, Qiang and de Rijke, Maarten and Cheng, Xiuzhen and Ren, Pengjie},
  booktitle={Proceedings of the ACM Web Conference 2023},
  pages={1283--1293},
  year={2023}
}

@inproceedings{zhang2022hierarchical,
  title={Hierarchical item inconsistency signal learning for sequence denoising in sequential recommendation},
  author={Zhang, Chi and Du, Yantong and Zhao, Xiangyu and Han, Qilong and Chen, Rui and Li, Li},
  booktitle={Proceedings of the 31st ACM international conference on information \& knowledge management},
  pages={2508--2518},
  year={2022}
}

@inproceedings{xie2022contrastive,
  title={Contrastive learning for sequential recommendation},
  author={Xie, Xu and Sun, Fei and Liu, Zhaoyang and Wu, Shiwen and Gao, Jinyang and Zhang, Jiandong and Ding, Bolin and Cui, Bin},
  booktitle={2022 IEEE 38th international conference on data engineering (ICDE)},
  pages={1259--1273},
  year={2022},
  organization={IEEE}
}

@inproceedings{wu2025empowering,
  title={Empowering Denoising Sequential Recommendation with Large Language Model Embeddings},
  author={Wu, Tongzhou and Wang, Yuhao and Wang, Maolin and Zhang, Chi and Zhao, Xiangyu},
  booktitle={Proceedings of the 34th ACM International Conference on Information and Knowledge Management},
  pages={3427--3437},
  year={2025}
}

@article{wang2025llm4dsr,
  title={Llm4dsr: Leveraging large language model for denoising sequential recommendation},
  author={Wang, Bohao and Liu, Feng and Zhang, Changwang and Chen, Jiawei and Wu, Yudi and Zhou, Sheng and Lou, Xingyu and Wang, Jun and Feng, Yan and Chen, Chun and others},
  journal={ACM Transactions on Information Systems},
  volume={44},
  number={1},
  pages={1--32},
  year={2025},
  publisher={ACM New York, NY}
}

@inproceedings{liu2025large,
  title={Large Language Model Enhanced Recommender Systems: Methods, Applications and Trends},
  author={Liu, Qidong and Zhao, Xiangyu and Wang, Yuhao and Wang, Yejing and Zhang, Zijian and Sun, Yuqi and Li, Xiang and Wang, Maolin and Jia, Pengyue and Chen, Chong and others},
  booktitle={Proceedings of the 31st ACM SIGKDD Conference on Knowledge Discovery and Data Mining V. 2},
  pages={6096--6106},
  year={2025}
}

@inproceedings{zhang2023denoising,
  title={Denoising and prompt-tuning for multi-behavior recommendation},
  author={Zhang, Chi and Chen, Rui and Zhao, Xiangyu and Han, Qilong and Li, Li},
  booktitle={Proceedings of the ACM web conference 2023},
  pages={1355--1363},
  year={2023}
}

@inproceedings{sun2025llm4rsr,
  title={LLM4RSR: Large Language Models as Data Correctors for Robust Sequential Recommendation},
  author={Sun, Yatong and Yang, Xiaochun and Sun, Zhu and Wang, Yan and Wang, Bin and Qu, Xinghua},
  booktitle={Proceedings of the AAAI Conference on Artificial Intelligence},
  volume={39},
  number={12},
  pages={12604--12612},
  year={2025}
}

@inproceedings{he2024double,
  title={Double correction framework for denoising recommendation},
  author={He, Zhuangzhuang and Wang, Yifan and Yang, Yonghui and Sun, Peijie and Wu, Le and Bai, Haoyue and Gong, Jinqi and Hong, Richang and Zhang, Min},
  booktitle={Proceedings of the 30th ACM SIGKDD Conference on Knowledge Discovery and Data Mining},
  pages={1062--1072},
  year={2024}
}

@inproceedings{zhang2025shapley,
  title={Shapley value-driven data pruning for recommender systems},
  author={Zhang, Yansen and Zhang, Xiaokun and Cui, Ziqiang and Ma, Chen},
  booktitle={Proceedings of the 31st ACM SIGKDD Conference on Knowledge Discovery and Data Mining V. 2},
  pages={3879--3888},
  year={2025}
}

@article{wu2024survey,
  title={A survey on large language models for recommendation},
  author={Wu, Likang and Zheng, Zhi and Qiu, Zhaopeng and Wang, Hao and Gu, Hongchao and Shen, Tingjia and Qin, Chuan and Zhu, Chen and Zhu, Hengshu and Liu, Qi and others},
  journal={World Wide Web},
  volume={27},
  number={5},
  pages={60},
  year={2024},
  publisher={Springer}
}

@article{ko2022survey,
  title={A survey of recommendation systems: recommendation models, techniques, and application fields},
  author={Ko, Hyeyoung and Lee, Suyeon and Park, Yoonseo and Choi, Anna},
  journal={Electronics},
  volume={11},
  number={1},
  pages={141},
  year={2022},
  publisher={MDPI}
}

@article{schafer2001commerce,
  title={E-commerce recommendation applications},
  author={Schafer, J Ben and Konstan, Joseph A and Riedl, John},
  journal={Data mining and knowledge discovery},
  volume={5},
  number={1},
  pages={115--153},
  year={2001},
  publisher={Springer}
}

@inproceedings{han2024efficient,
  title={Efficient noise-decoupling for multi-behavior sequential recommendation},
  author={Han, Yongqiang and Wang, Hao and Wang, Kefan and Wu, Likang and Li, Zhi and Guo, Wei and Liu, Yong and Lian, Defu and Chen, Enhong},
  booktitle={Proceedings of the ACM Web Conference 2024},
  pages={3297--3306},
  year={2024}
}

@inproceedings{zhang2024ssdrec,
  title={Ssdrec: Self-augmented sequence denoising for sequential recommendation},
  author={Zhang, Chi and Han, Qilong and Chen, Rui and Zhao, Xiangyu and Tang, Peng and Song, Hongtao},
  booktitle={2024 IEEE 40th International Conference on Data Engineering (ICDE)},
  pages={803--815},
  year={2024},
  organization={IEEE}
}

@inproceedings{chen2022denoising,
  title={Denoising self-attentive sequential recommendation},
  author={Chen, Huiyuan and Lin, Yusan and Pan, Menghai and Wang, Lan and Yeh, Chin-Chia Michael and Li, Xiaoting and Zheng, Yan and Wang, Fei and Yang, Hao},
  booktitle={Proceedings of the 16th ACM conference on recommender systems},
  pages={92--101},
  year={2022}
}

@inproceedings{gao2022sgdl,
  title={Self-guided learning to denoise for robust recommendation},
  author={Gao, Yunjun and Du, Yuntao and Hu, Yujia and Chen, Lu and Zhu, Xinjun and Fang, Ziquan and Zheng, Baihua},
  booktitle={Proceedings of the 45th international ACM SIGIR conference on research and development in information retrieval},
  pages={1412--1422},
  year={2022}
}

@inproceedings{hu2024semanticembedding,
  title={Enhancing sequential recommendation via llm-based semantic embedding learning},
  author={Hu, Jun and Xia, Wenwen and Zhang, Xiaolu and Fu, Chilin and Wu, Weichang and Huan, Zhaoxin and Li, Ang and Tang, Zuoli and Zhou, Jun},
  booktitle={Companion Proceedings of the ACM Web Conference 2024},
  pages={103--111},
  year={2024}
}

@inproceedings{li2024large,
  title={Large language models for generative recommendation: A survey and visionary discussions},
  author={Li, Lei and Zhang, Yongfeng and Liu, Dugang and Chen, Li},
  booktitle={Proceedings of the 2024 Joint International Conference on Computational Linguistics, Language Resources and Evaluation (LREC-COLING 2024)},
  pages={10146--10159},
  year={2024}
}

@article{zhang2025robust,
  title={Robust recommender system: a survey and future directions},
  author={Zhang, Kaike and Cao, Qi and Sun, Fei and Wu, Yunfan and Tao, Shuchang and Shen, Huawei and Cheng, Xueqi},
  journal={ACM Computing Surveys},
  volume={58},
  number={1},
  pages={1--38},
  year={2025},
  publisher={ACM New York, NY}
}

@inproceedings{sun2021does,
  title={Does Every Data Instance Matter? Enhancing Sequential Recommendation by Eliminating Unreliable Data.},
  author={Sun, Yatong and Wang, Bin and Sun, Zhu and Yang, Xiaochun},
  booktitle={IJCAI},
  pages={1579--1585},
  year={2021}
}

@article{song2024large,
  title={Large language model enhanced hard sample identification for denoising recommendation},
  author={Song, Tianrui and Chao, Wenshuo and Liu, Hao},
  journal={arXiv preprint arXiv:2409.10343},
  year={2024}
}

@article{chen2023bias,
  title={Bias and debias in recommender system: A survey and future directions},
  author={Chen, Jiawei and Dong, Hande and Wang, Xiang and Feng, Fuli and Wang, Meng and He, Xiangnan},
  journal={ACM Transactions on Information Systems},
  volume={41},
  number={3},
  pages={1--39},
  year={2023},
  publisher={ACM New York, NY}
}

@inproceedings{wang2023efficient,
  title={Efficient bi-level optimization for recommendation denoising},
  author={Wang, Zongwei and Gao, Min and Li, Wentao and Yu, Junliang and Guo, Linxin and Yin, Hongzhi},
  booktitle={Proceedings of the 29th ACM SIGKDD conference on knowledge discovery and data mining},
  pages={2502--2511},
  year={2023}
}

@article{vaswani2017attention,
  title={Attention is all you need},
  author={Vaswani, Ashish and Shazeer, Noam and Parmar, Niki and Uszkoreit, Jakob and Jones, Llion and Gomez, Aidan N and Kaiser, {\L}ukasz and Polosukhin, Illia},
  journal={Advances in neural information processing systems},
  volume={30},
  year={2017}
}

@article{wang2025mllm4rec,
  title={Mllm4rec: multimodal information enhancing llm for sequential recommendation},
  author={Wang, Yuxiang and Shi, Xin and Zhao, Xueqing},
  journal={Journal of Intelligent Information Systems},
  volume={63},
  number={3},
  pages={745--761},
  year={2025},
  publisher={Springer}
}

@article{boz2025improving,
  title={Improving sequential recommendations with llms},
  author={Boz, Artun and Zorgdrager, Wouter and Kotti, Zoe and Harte, Jesse and Louridas, Panos and Karakoidas, Vassilios and Jannach, Dietmar and Fragkoulis, Marios},
  journal={ACM Transactions on Recommender Systems},
  volume={4},
  number={2},
  pages={1--35},
  year={2025},
  publisher={ACM New York, NY}
}

@article{zhao2023survey,
  title={A survey of large language models},
  author={Zhao, Wayne Xin and Zhou, Kun and Li, Junyi and Tang, Tianyi and Wang, Xiaolei and Hou, Yupeng and Min, Yingqian and Zhang, Beichen and Zhang, Junjie and Dong, Zican and others},
  journal={arXiv preprint arXiv:2303.18223},
  volume={1},
  number={2},
  year={2023}
}

@article{kasneci2023chatgpt,
  title={ChatGPT for good? On opportunities and challenges of large language models for education},
  author={Kasneci, Enkelejda and Se{\ss}ler, Kathrin and K{\"u}chemann, Stefan and Bannert, Maria and Dementieva, Daryna and Fischer, Frank and Gasser, Urs and Groh, Georg and G{\"u}nnemann, Stephan and H{\"u}llermeier, Eyke and others},
  journal={Learning and individual differences},
  volume={103},
  pages={102274},
  year={2023},
  publisher={Elsevier}
}

@article{ji2023survey,
  title={Survey of hallucination in natural language generation},
  author={Ji, Ziwei and Lee, Nayeon and Frieske, Rita and Yu, Tiezheng and Su, Dan and Xu, Yan and Ishii, Etsuko and Bang, Ye Jin and Madotto, Andrea and Fung, Pascale},
  journal={ACM computing surveys},
  volume={55},
  number={12},
  pages={1--38},
  year={2023},
  publisher={ACM New York, NY}
}

@inproceedings{he2016ups,
  title={Ups and downs: Modeling the visual evolution of fashion trends with one-class collaborative filtering},
  author={He, Ruining and McAuley, Julian},
  booktitle={proceedings of the 25th international conference on world wide web},
  pages={507--517},
  year={2016}
}

@inproceedings{zhou2022filter,
  title={Filter-enhanced MLP is all you need for sequential recommendation},
  author={Zhou, Kun and Yu, Hui and Zhao, Wayne Xin and Wen, Ji-Rong},
  booktitle={Proceedings of the ACM web conference 2022},
  pages={2388--2399},
  year={2022}
}

@article{touvron2023llama,
  title={Llama: Open and efficient foundation language models},
  author={Touvron, Hugo and Lavril, Thibaut and Izacard, Gautier and Martinet, Xavier and Lachaux, Marie-Anne and Lacroix, Timoth{\'e}e and Rozi{\`e}re, Baptiste and Goyal, Naman and Hambro, Eric and Azhar, Faisal and others},
  journal={arXiv preprint arXiv:2302.13971},
  year={2023}
}

@article{hu2022lora,
  title={Lora: Low-rank adaptation of large language models.},
  author={Hu, Edward J and Shen, Yelong and Wallis, Phillip and Allen-Zhu, Zeyuan and Li, Yuanzhi and Wang, Shean and Wang, Lu and Chen, Weizhu and others},
  journal={ICLR},
  volume={1},
  number={2},
  pages={3},
  year={2022}
}

@article{bai2023qwen,
  title={Qwen technical report},
  author={Bai, Jinze and Bai, Shuai and Chu, Yunfei and Cui, Zeyu and Dang, Kai and Deng, Xiaodong and Fan, Yang and Ge, Wenbin and Han, Yu and Huang, Fei and others},
  journal={arXiv preprint arXiv:2309.16609},
  year={2023}
}

\appendix

\section{Reproducibility Details of Disagreement-guided Calibration}
\label{app:calibration}
We run calibration every $R$ epochs for a total of $\lfloor E/R \rfloor$ rounds.
For each queried sequence $x\in\mathcal{P}$, we refresh the semantic prior only on hard positions $i\in\mathcal{H}(x)$ using the same fixed-prefix next-token scoring protocol as in Sec.~\ref{sec:method_semantic_prior}.
Concretely, for each $i\in\mathcal{H}(x)$ we compute an updated score $s_i\in[0,1]$ from the next-token probability at the fixed response prefix, and set $p_{\mathrm{sem},x}[i]\leftarrow s_i$.
The prompt template and serialization schema are provided in Appendix~\ref{app:prompt}.

\subsection{Prompt Template and Output Schema}
\label{app:prompt}

We query the LLM only for selected sequences $x\in\mathcal{P}$ and their hard positions $\mathcal{H}(x)$.
The query is serialized with a fixed prompt template (Table~\ref{tab:prompt_spec}), and the response prefix \texttt{Suspicious items:} is preserved for consistent next-token scoring.

\begin{table}[h]
\centering
\caption{Prompt specification for semantic-prior refresh. The evidence log $\mathcal{E}(x)$ is injected into the input, while the fixed output prefix \texttt{Suspicious items:} is preserved to enable the same next-token scoring protocol as in Sec.~\ref{sec:method_semantic_prior}.}
\small
\setlength{\tabcolsep}{6pt}
\renewcommand{\arraystretch}{1.10}
\begin{tabular}{p{0.18\linewidth} p{0.7\linewidth}}
\hline
Field & Template \\
\hline

\texttt{instruction} &
{\ttfamily
You are a recommender systems researcher familiar with noisy user behavior logs and item title semantics.
You will be given a user behavior sequence and an evidence log that records which positions are contentious across refresh rounds.
Focus on the listed hard positions and identify which of them are suspicious noise items under the context.
} \\

\texttt{input} &
{\ttfamily
User behavior sequence (chronological):\newline
1. <title/text> \;|\; 2. <title/text> \;|\; ... \;|\; n. <title/text>\newline
Hard positions (1-indexed): <pos\_1>, <pos\_2>, ..., <pos\_K>\newline
Evidence log (per hard position, last $L$ traces):\newline
pos <pos\_j>: (p\_sem, p\_mod, d) at recent rounds: <(.,.,.), ..., (.,.,.)>\newline
(Optional) pool summary: <text>
} \\

\texttt{output} &
{\ttfamily
Suspicious items:\newline
\{ <title/text at pos>, <title/text at pos>, ... \}\newline
If none:\; \{\}
} \\

\hline
\end{tabular}
\label{tab:prompt_spec}
\end{table}

\begin{algorithm}[h]
\caption{Disagreement-guided Calibration}
\label{alg:calibration}
\DontPrintSemicolon
\SetKwInOut{Input}{Input}
\SetKwInOut{Output}{Output}
\Input{Training set $\mathcal{D}$; total epochs $E$; refresh interval $R$; pool size $M$; hard spots per sequence $K$; evidence window length $L$; LLM scorer $\mathcal{M}_{\phi}$.}
\Input{Semantic priors $\{p_{\mathrm{sem},x}[i]\}$; model risks $\{p_{\mathrm{mod},x}[i]\}$; disagreement $d_x[i]=|p_{\mathrm{sem},x}[i]-p_{\mathrm{mod},x}[i]|$ for $i\in\mathcal{V}(x)$.}
\Output{Updated semantic priors $\{p_{\mathrm{sem},x}[i]\}$ on queried positions.}

\For{$e \leftarrow R, 2R, \ldots, \lfloor E/R \rfloor R$}{
  \ForEach{$x \in \mathcal{D}$}{
    $U(x) \leftarrow \frac{1}{|\mathcal{V}(x)|}\sum_{i \in \mathcal{V}(x)} d_x[i]$\;
  }
  $\mathcal{P} \leftarrow \text{TopMSequences}(U(\cdot); \mathcal{D}, M)$\;

  \ForEach{$x \in \mathcal{P}$}{
    $\mathcal{H}(x) \leftarrow \text{TopKPositions}(\{d_x[i]\}_{i\in\mathcal{V}(x)}, K)$\;

    \ForEach{$i \in \mathcal{H}(x)$}{
      append $(p_{\mathrm{sem},x}[i],\,p_{\mathrm{mod},x}[i],\,d_x[i])$ to $\mathcal{E}(x,i)$\;
      keep the most recent $L$ entries in $\mathcal{E}(x,i)$\;
    }

    build prompt $\mathcal{Q}(x)$ from $\{\mathcal{E}(x,i)\}_{i\in \mathcal{H}(x)}$ and item texts\;
    run $\mathcal{M}_{\phi}$ with fixed output prefix to obtain updated scores $\{s_i\in[0,1]\}_{i\in\mathcal{H}(x)}$\;
    \ForEach{$i \in \mathcal{H}(x)$}{
      $p_{\mathrm{sem},x}[i] \leftarrow s_i$\;
    }
  }
}
\end{algorithm}


\begin{table}[t]
\centering
\caption{Traction-based down-weighting results on the Office dataset.}
\label{tab:traction}
\small
\setlength{\tabcolsep}{4.2pt}
\renewcommand{\arraystretch}{1.15}
\begin{tabular}{lccc}
\toprule
Metric & Original & Adaptive & Related \\
\midrule
HR@20   & 0.04613 & \textbf{0.04806} & 0.04735 \\
NDCG@20 & 0.01671 & \textbf{0.01741} & 0.01733 \\
\bottomrule
\end{tabular}
\end{table}

\begin{figure}[t]
  \centering
  \begin{minipage}[t]{0.49\linewidth}
    \centering
    \includegraphics[width=\linewidth]{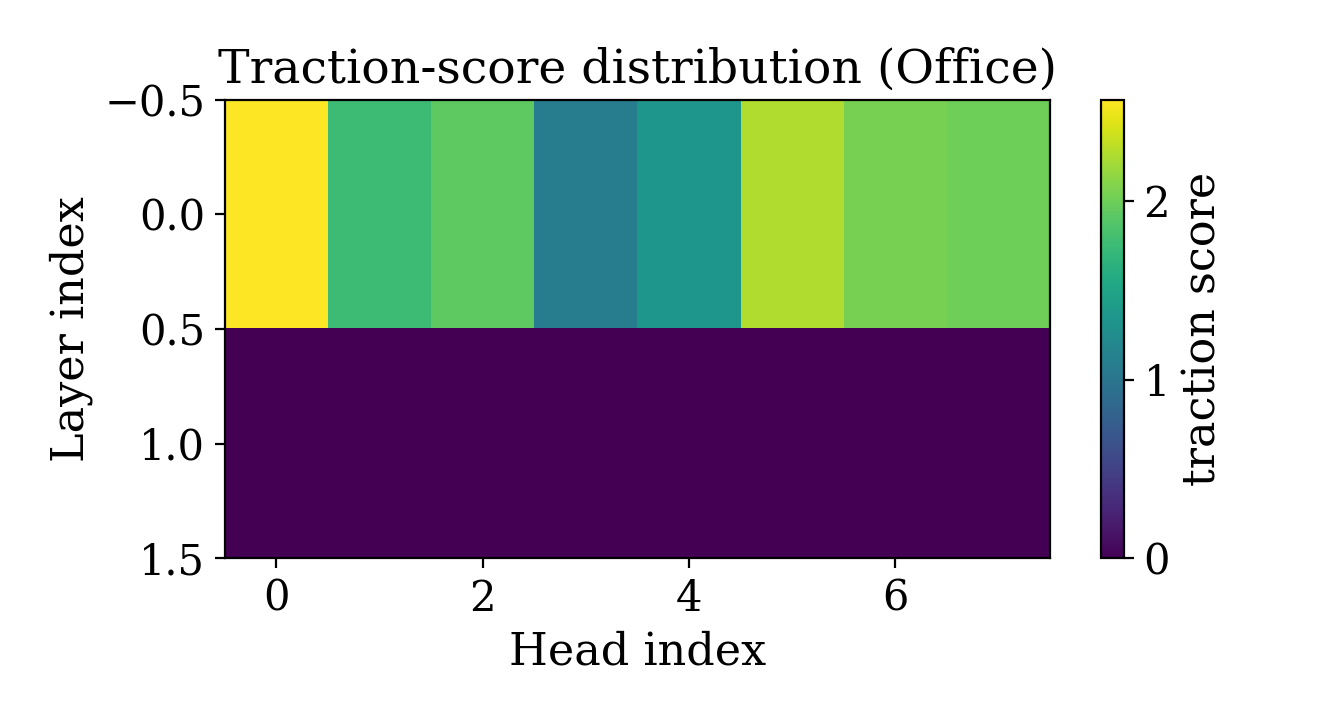}
    \caption*{\small (a) Head-wise traction score.}
  \end{minipage}\hfill
  \begin{minipage}[t]{0.49\linewidth}
    \centering
    \includegraphics[width=\linewidth]{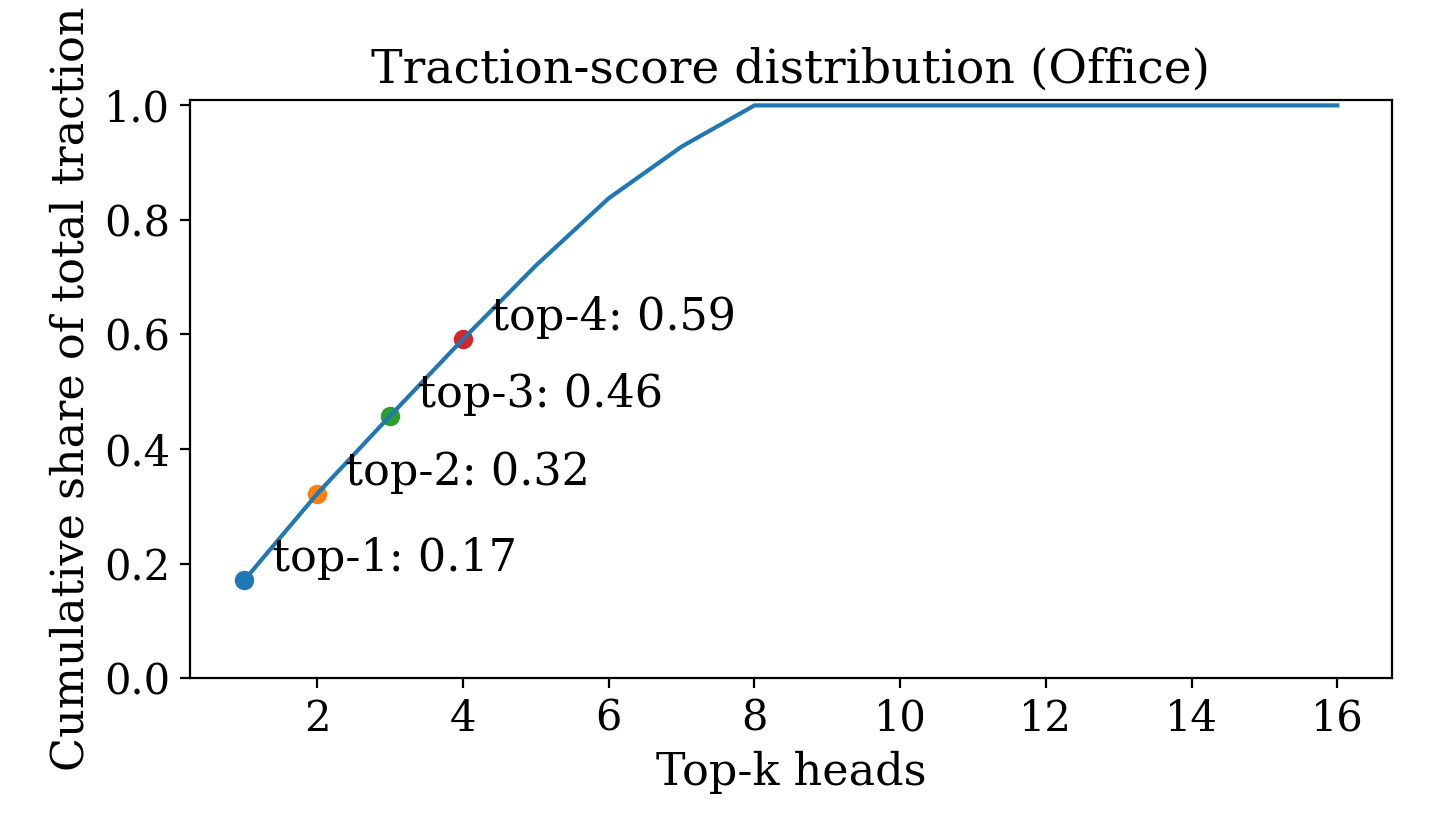}
    \caption*{\small (b) Cumulative traction mass.}
  \end{minipage}
  \caption{Traction-score diagnostics on Office with SASRec. (a) Traction scores are highly non-uniform and mainly localized in the first block. (b) The top-4 heads account for 59\% of the total traction mass, motivating $k{=}4$ for down-weighting.}
  \label{fig:traction_diagnosis}
\end{figure}

\section{Additional Results for Traction Analysis}
\label{app:traction}

We provide additional results for the traction-based experiment described in Sec.~2.3.
We use a SASRec backbone with $L{=}2$ Transformer blocks and $H{=}8$ attention heads per block (i.e., $LH{=}16$ head instances in total) on the Office dataset.
For each head $(l,h)$, we measure its injection sensitivity by the expected absolute change of the query--key logit at the decision position:
\begin{equation}
g_{l,h}=\mathbb{E}\big[\,|\ell^{\mathrm{inj}}_{l,h}-\ell^{\mathrm{orig}}_{l,h}|\,\big],
\end{equation}
where $\ell_{l,h}(t,j)=\langle Q_{l,h}(t),K_{l,h}(j)\rangle/\sqrt{d}$ denotes the pre-softmax attention logit from the last query position $t$ to the injected key position $j$.
We rank all heads by $g_{l,h}$ and select the top-$k$ heads as traction heads.

Figure~\ref{fig:traction_diagnosis} empirically confirms that the traction-head phenomenon is present in a trained SASRec model on Office.
The heatmap shows a highly non-uniform head-wise response to injections, with the sensitivity mainly localized in the first Transformer block rather than being evenly distributed across heads and layers.
The cumulative curve further indicates a compact concentration: the top-4 heads already account for 59\% of the total traction score, which motivates setting $k{=}4$.

At inference time, we down-weight the selected traction heads by head-wise dynamic coefficients $\gamma_{l,h}\in(0,1)$ that monotonically decrease with $g_{l,h}$, while leaving other heads unchanged. Table~\ref{tab:traction} reports the original inference and traction-head down-weighting (Adaptive), together with a semantically-related injection control (Related), which identifies responsive heads via semantically related injections as described in the main text, while using the same down-weighting scheme.

\begin{table}[t]
\centering
\small
\setlength{\tabcolsep}{5pt}
\renewcommand{\arraystretch}{1.1}
\caption{Dataset statistics.}
\label{tab:dataset_stats}
\begin{tabular}{lrrrr}
\toprule
Dataset & \#Users & \#Items & \#Interactions & Density \\
\midrule
Movie  & 6884  & 5034  & 1070872 & 3.09e-02 \\
Toys   & 14117 & 11577 & 368093  & 2.25e-03 \\
Office & 3718  & 2405  & 158482  & 1.77e-02 \\
\bottomrule
\end{tabular}
\end{table}

\section{Dataset Statistics}
\label{app:dataset}

Table~\ref{tab:dataset_stats} reports basic statistics of the datasets used in our experiments.
All datasets are derived from implicit feedback logs and organized as user-level interaction sequences.
We report the number of users, items, and interactions, and compute the data density as the ratio between observed interactions and the total possible user--item pairs.
Statistics are computed on the training split, excluding padding items.

\end{document}